\pdfoutput=1

\documentclass[sigconf, nonacm, pdfa]{acmart}
\usepackage[utf8]{inputenc}

\newcommand\vldbdoi{10.14778/3717755.3717759}
\newcommand\vldbpages{970 - 985}
\newcommand\vldbvolume{18}
\newcommand\vldbissue{4}
\newcommand\vldbyear{2024}
\newcommand\vldbauthors{\authors}
\newcommand\vldbtitle{\shorttitle} 
\newcommand\vldbavailabilityurl{https://github.com/illinoisdata/kishu-vldb}
\newcommand\vldbpagestyle{empty}

\usepackage{bm}
\usepackage[linesnumbered,ruled]{algorithm2e}
\usepackage{algorithmic}
\usepackage{hyperref}
\usepackage[capitalise]{cleveref}
\crefname{section}{§}{§§}
\Crefname{section}{§}{§§}
\crefformat{section}{\S#2#1#3}
\crefname{figure}{Fig}{Figs}
\Crefname{figure}{Fig}{Figs}
\crefname{problem}{Prob}{Probs}
\Crefname{problem}{Prob}{Probs}
\usepackage{subfiles} 
\usepackage{multirow}
\usepackage{tabularx}
\usepackage{xspace}
\usepackage{subcaption}
\usepackage{pgfplots}
\usepackage{pgfplotstable}
\usepackage{xcolor}
\usepackage{ifthen}
\usepackage[shortlabels]{enumitem}
\setlist[enumerate]{leftmargin=1cm,topsep=0.5mm}
\setlist[itemize]{leftmargin=0.5cm,topsep=0.5mm}

\usepackage{tikz}
\usetikzlibrary{calc}
\usetikzlibrary{fit}
\usetikzlibrary{patterns}
\usetikzlibrary{decorations.pathreplacing}
\usetikzlibrary{arrows,calc,arrows.meta}
\usetikzlibrary{matrix,positioning}
\usetikzlibrary{shapes.geometric}

\definecolor{Ours0Color}{HTML}{ABDDA4}
\definecolor{Ours16Color}{HTML}{72C166}
\definecolor{Ours32Color}{HTML}{38A528}
\definecolor{PrestoColor}{HTML}{999999}
\definecolor{PostgresColor}{HTML}{6D6D6D}
\definecolor{GreenColor}{HTML}{38A528}
\definecolor{YellowColor}{HTML}{ffb570}
\definecolor{BlueColor}{HTML}{7081ff}
\definecolor{PinkColor}{HTML}{ffb0c2}

\definecolor{ComputeColor}{HTML}{ffb0c2}
\definecolor{ReadColor}{HTML}{cf3457}
\definecolor{WriteColor}{HTML}{ffb570}

\definecolor{vintagegreen}{HTML}{ABDDA4}
\definecolor{OursColor}{HTML}{38A528}
\definecolor{GreedyColor}{HTML}{7081ff}
\definecolor{RandomColor}{HTML}{ffb570}
\definecolor{NoneColor}{HTML}{6D6D6D}

\definecolor{Redborder}{HTML}{805861}
\definecolor{Greenborder}{HTML}{384180}
\definecolor{Blueborder}{HTML}{566F52}
\definecolor{Greyborder}{HTML}{4D4D4D}
\definecolor{Lightgrey}{HTML}{dadada}
\definecolor{ExampleColor1}{HTML}{7081ff}
\definecolor{ExampleColor2}{HTML}{ffb0c2}
\definecolor{Lightred}{HTML}{ffb09c}
\definecolor{Lightblue}{HTML}{b8e2f2}
\definecolor{FlagColor}{HTML}{CCCCCC}

\definecolor{vintageblue}{HTML}{7081ff}
\definecolor{vintagered}{HTML}{ffb0c2}

\definecolor{NoOptColor}{HTML}{264653}
\definecolor{LRUColor}{HTML}{777777}
\definecolor{RandomColor}{HTML}{2a9d8f}
\definecolor{GreedyColor}{HTML}{e9c46a}
\definecolor{HeuristicColor}{HTML}{f4a261}
\definecolor{SCColor}{HTML}{e76f51}
\definecolor{AllColor}{HTML}{CCCCCC}

\definecolor{SAColor}{HTML}{ffb0c2}
\definecolor{SeparatorColor}{HTML}{9b5de5}
\definecolor{BlueColor}{HTML}{0081a7}

\newcommand{\midsepremove}{\aboverulesep = 0.3mm \belowrulesep = 0.3mm}
    \midsepremove
    \newcommand{\midsepdefault}{\aboverulesep = 0.605mm \belowrulesep = 0.984mm}
    \midsepdefault

\usepackage{amsthm}
\theoremstyle{definition}
\newtheorem{lemma}{Lemma}
\newtheorem{definition}{Definition}

\newcommand{\system}{{\sf Kishu}\xspace}
\newcommand{\systembf}{{\sffamily\bfseries Kishu}\xspace}
\newcommand{\systemnosf}{Kishu\xspace}
\newcommand{\systembad}{{\sf AblatedKishu (Check all)}\xspace}

\newcommand{\elasticnotebook}{{\sf ElasticNotebook}\xspace}
\newcommand{\ipyflow}{{\sf IPyFlow}\xspace}

\newcommand{\criu}{{\sf CRIU}\xspace}
\newcommand{\criuincremental}{{\sf CRIU-Incremental}\xspace}
\newcommand{\detreplay}{{\sf Kishu+Det-replay}\xspace}
\newcommand{\vargraph}{\textsf{VarGraph}\xspace}
\newcommand{\vargraphs}{\textsf{VarGraphs}\xspace}

\newcommand{\dumpsession}{{\sf DumpSession}\xspace}

\definecolor{vintagegray}{HTML}{e5e5e5}
\newtheoremstyle{abcd}
  {}
  {}
  {\itshape}
  {}
  {\bfseries}
  {.}
  {.5em}
  {}
\theoremstyle{abcd}
\usepackage{pgfplots}
\setenumerate{label={\arabic*.},noitemsep,topsep=2pt,leftmargin=14pt}

\makeatletter
\newcommand\resetstackedplots{
\makeatletter
\pgfplots@stacked@isfirstplottrue
\makeatother

\addplot [forget plot,draw=none] coordinates{(1,0) (2,0) (3,0)};
}

\makeatletter
\newcommand{\problemtitle}[1]{\gdef\@problemtitle{#1}}
\newcommand{\probleminput}[1]{\gdef\@probleminput{#1}}
\newcommand{\problemoutput}[1]{\gdef\@problemoutput{#1}}
\newcommand{\problemobjective}[1]{\gdef\@problemobjective{#1}}
\newcommand{\problemconstraint}[1]{\gdef\@problemconstraint{#1}}
\NewEnviron{myproblem}{
  \problemtitle{}\probleminput{}\problemquestion{}
  \BODY
  \par\addvspace{.5\baselineskip}
  \noindent
  \begin{tabularx}{\textwidth}{@{\hspace{\parindent}} l X c}
    \multicolumn{2}{@{\hspace{\parindent}}l}{\@problemtitle} \\
    \textbf{Input:} & \@probleminput \\
    \textbf{Output:} & \@problemoutput
    \textbf{Objective Function:} & \@problemobjective
    \textbf{Constraint:} & \@problemconstraint
  \end{tabularx}
  \par\addvspace{.5\baselineskip}
}
\makeatother

\begin{document}
\title[Kishu: Time-Traveling for Computational Notebooks]{Kishu: Time-Traveling for Computational Notebooks}

\author[Zhaoheng Li, Supawit Chockchowwat, Ribhav Sahu, Areet Sheth, Yongjoo Park]{Zhaoheng Li, Supawit Chockchowwat, Ribhav Sahu, Areet Sheth, Yongjoo Park}
\affiliation{%
  \institution{University of Illinois at Urbana-Champaign}
}
\email{{zl20,supawit2,ribhav2,assheth2,yongjoo}@illinois.edu}








\begin{abstract} Computational notebooks (e.g., Jupyter, Google Colab) are widely used by data scientists. A key feature of notebooks is the interactive computing model of iteratively executing \emph{cells} (i.e., a set of statements) and observing the result (e.g., model or plot).
Unfortunately, existing notebook systems do not offer \emph{time-traveling to past states}: when the user executes a cell, the notebook \emph{session state} consisting of user-defined variables can be \emph{irreversibly modified}---e.g., the user cannot 'un-drop' a dataframe column.
This is because, unlike DBMS, existing notebook systems do not keep track of the session state.
Existing techniques for checkpointing and restoring session states, such as OS-level memory snapshot or application-level session dump, are insufficient:
checkpointing can incur prohibitive storage costs and may fail, while restoration can only be inefficiently performed from scratch by fully loading checkpoint files.

In this paper, we introduce a new notebook system, \system, that offers time-traveling to and from arbitrary notebook states using an efficient and fault-tolerant incremental checkpoint and checkout mechanism.
\system creates incremental checkpoints that are small and correctly preserve complex inter-variable dependencies at a novel \emph{Co-variable} granularity.
Then, to return to a previous state, \system accurately identifies the \emph{state difference} between the current and target states to perform incremental checkout at sub-second latency with minimal data loading.
\system is compatible with 146 object classes from popular data science libraries (e.g., Ray, Spark, PyTorch), and reduces checkpoint size and checkout time by up to 4.55$\times$ and 9.02$\times$, respectively, on a variety of notebooks.

\end{abstract}

\maketitle


\pagestyle{\vldbpagestyle}
\begingroup\small\noindent\raggedright\textbf{PVLDB Reference Format:}\\
\vldbauthors. \vldbtitle. PVLDB, \vldbvolume(\vldbissue): \vldbpages, \vldbyear.\\
\href{https://doi.org/\vldbdoi}{doi:\vldbdoi}
\endgroup
\begingroup
\renewcommand\thefootnote{}\footnote{\noindent
This work is licensed under the Creative Commons BY-NC-ND 4.0 International License. Visit \url{https://creativecommons.org/licenses/by-nc-nd/4.0/} to view a copy of this license. For any use beyond those covered by this license, obtain permission by emailing \href{mailto:info@vldb.org}{info@vldb.org}. Copyright is held by the owner/author(s). Publication rights licensed to the VLDB Endowment. \\
\raggedright Proceedings of the VLDB Endowment, Vol. \vldbvolume, No. \vldbissue\ %
ISSN 2150-8097. \\
\href{https://doi.org/\vldbdoi}{doi:\vldbdoi} \\
}\addtocounter{footnote}{-1}\endgroup

\ifdefempty{\vldbavailabilityurl}{}{
\vspace{.3cm}
\begingroup\small\noindent\raggedright\textbf{PVLDB Artifact Availability:}\\
The source code, data, and/or other artifacts have been made available at \url{\vldbavailabilityurl}.
\endgroup
}

\section{Introduction}
\label{sec:intro}

\begin{figure}[t]

\tikzset{
codenode/.style={
    draw=black,minimum width=16mm,
    font=\small\ttfamily,
},
cellnode/.style={
    font=\small\sffamily,
    anchor=south west,inner ysep=0,
},
technode/.style={
    draw=black,text width=34mm,
    align=left,
    font=\small\sffamily,
    inner ysep=0.5mm
},
mylabel/.style={
    font=\footnotesize\sffamily\bfseries,
    align=center,
},
mylabel2/.style={
    font=\footnotesize\sffamily,
    align=center,
},
mycomponent/.style={
    semithick, rounded corners=0.5mm,
}
}

\centering
\begin{tikzpicture}

\def\g{0.5}


\node[draw=black,thick, fill = Lightgrey, minimum width=24mm, minimum height=17mm] (B) at (0, 0) {};
\node[anchor=south,font=\small\bfseries] at (B.north) 
    {Notebook Interface};

\node(cell2) [draw=black, anchor=north east, minimum width = 19mm, minimum height=6mm, fill = white]
at ($(B.north east) + (-0.1, -0.1)$) {};
\node(codetxt2) [anchor=west, minimum height=3mm, inner sep = 0.6mm, align=left]
at ($(cell2.west) + (0.05, 0)$) {\footnotesize \texttt{import pandas}\\[-0.35em]\footnotesize \texttt{df=pd.read\_csv}};

\node(cell3) [draw=black, anchor=north east, minimum width = 19mm, minimum height=3.5mm, fill = white]
at ($(cell2.south east) + (0, -0.1)$) {};
\node(codetxt3) [anchor=west, minimum height=3mm, inner sep = 0.6mm, align=left]
at ($(cell3.west) + (0.05, 0)$) {\footnotesize \texttt{df.head}};

\node(cell4) [anchor=north west, minimum width = 14mm, minimum height=4.5mm, fill = white]
at ($(cell3.south west) + (0, -0.05)$) {};

\node[anchor=east,font=\footnotesize] at ($(cell2.north west) + (0.1, -0.15)$) 
    {[1]};
\node[anchor=east,font=\footnotesize] at ($(cell3.north west) + (0.1, -0.15)$) 
    {[2]};

\node[anchor=west,font=\footnotesize] at ($(cell3.south west) + (0, -0.15)$) 
    {col1};
\node[anchor=west,font=\footnotesize] at ($(cell3.south west) + (0.55, -0.15)$) 
    {col2};
\node[anchor=west,font=\footnotesize] at ($(cell3.south west) + (0, -0.4)$) 
    {1};
\node[anchor=west,font=\footnotesize] at ($(cell3.south west) + (0.55, -0.4)$) 
    {hello};
    
\draw[-] 
($(cell3.south west) + (0, -0.27)$) --
($(cell3.south west) + (1.4, -0.27)$);


\node[draw=black,thick,anchor=west,
    minimum width=8mm,minimum height=16mm,align=center,mycomponent,
    font=\footnotesize\sffamily] 
    (B3) at ($(B.east)+(\g,0)$) {\baselineskip=0pt
        Python/ \\
        R/LLVM \\
        Kernel
    };

\node[draw=black,thick,anchor=west,
    minimum width=38mm,minimum height=17mm, mycomponent] 
    (B2) at ($(B3.east)+(\g,0)$) {};
\node[anchor=south,font=\small\bfseries] 
    at ($(B2.north)+(0,-0.08)$)
    {Our System (\systembf)};
\node[anchor=north west,technode,minimum height=7mm] 
    (T1) at ($(B2.north west)+(0.1,-0.1)$)
    {\baselineskip=-0.2pt \textbf{Incremental Checkpoint}\\[-0.3em] via fine-grained deltas (\cref{sec:checkpoint})};
\node[anchor=north west,technode,minimum height=7mm] 
    (T2) at ($(T1.south west)+(0,-0.1)$)
    {\baselineskip=-0.2pt \textbf{Incremental Checkout} via\\[-0.35em] minimal data loading (\cref{sec:restore})};

\draw[-latex, ultra thick] 
    ($(B.east)+(0.05,0.2)$) -- ($(B3.west)+(-0.05,0.2)$);
\draw[latex-, ultra thick] 
    ($(B.east)+(0.05,-0.2)$) -- ($(B3.west)+(-0.05,-0.2)$);
\draw[-latex, ultra thick] 
    ($(B3.east)+(0.05,0.2)$) -- ($(B2.west)+(-0.05,0.2)$);
\draw[latex-, ultra thick] 
    ($(B3.east)+(0.05,-0.2)$) -- ($(B2.west)+(-0.05,-0.2)$);

\end{tikzpicture}

\caption{Our system (attached to the kernel, right) enables time-traveling to and from arbitrary notebook states.}
\end{figure}


\begin{table*}[t]
\centering
\caption{Comparison between \systembf and 
    other possible approaches for time-traveling between session states}
\label{tbl:existing_work}
\addtolength{\tabcolsep}{-2pt} 
\small
\begin{tabular}{lll}
\toprule
\textbf{Approach}              &  \textbf{Mechanism}           \\ \midrule
Session Migration tools~\cite{dill, lielasticnotebook}  & Replicates individual session states (no incremental checkpoint and restore)\\

OS-level Checkpoint tools~\cite{criu, juric2021checkpoint, ansel2009dmtcp, garg2018crum, jain2020crac} & Incrementally saves dirty memory pages (no incremental restore, fails with multiprocessing) \\

Notebook code versioning~\cite{brachmann2019data, guo2012burrito, jupytercheckpoint} & Versions notebook cell code and outputs for easier rerunning (but no directly restoring to a state) \\

   
 
PITR/time-travel in DBMS/HPC~\cite{lin2020incremental, keller2019application, schule2019versioning, antonopoulos2019constant, mysql} 
  
  & Incrementally checkpoints tables/KV-stores (We handle arbitrarily complex, interdependent objects)\\    

\textbf{Ours (\system)} 
& \textbf{Efficient incremental checkpoint and checkout, generalizable to almost all session states} \\
\bottomrule
\end{tabular}
\addtolength{\tabcolsep}{2pt} 
\end{table*}


\noindent
Computational notebooks (e.g., Jupyter~\cite{jupyter, ipython}, Rstudio~\cite{rstudio}) are widely used by data scientists~\cite{ormond2018acm, perkel2018jupyter}. A key feature of the notebook workflow is iterative code execution and result observation~\cite{amershi2019software, chattopadhyay2020s}, which is highly compatible with the incremental nature of data science tasks, such as interactive tutorials~\cite{johnson2020benefits},
data exploration~\cite{crotty2015vizdom, zgraggen2014panoramicdata, dunne2012graphtrail}, 
visualization~\cite{eichmann2020idebench}, and model tuning~\cite{wagenmakers2004aic, bergstra2012random}.
This iterative workflow is enabled by notebooks systems being \textit{stateful}---to do work, users would start a \textit{session}, then as users execute code in the notebook system, the results are held in the \textit{session state} as user-defined variables (e.g., loaded datasets, fitted models).


\paragraph{Limitation: no Time-Traveling for Notebooks} 
Oftentimes, during a workflow, users would like to revert changes made to the session state (i.e., \textit{time-travel}), such as to undo a modification (e.g., restore a dropped column of a dataframe~\cite{dropcol}), restoring an overwritten variable~\cite{jupyterundo}, or perform reverse debugging~\cite{brachmann2020your}.
Unfortunately, unlike program debuggers (e.g., gdb)~\cite{phang2013expositor, barr2014tardis, gdbreverse}, relational databases (e.g., PITR in PostgreSQL and MySQL~\cite{postgresqlpitr, mysqlpitr}) or interactive data systems~\cite{dunne2012graphtrail, crotty2015vizdom, kraska2021northstar} which support time-traveling to past program states, existing notebook systems do not natively keep track of past session states: cell executions \textit{cannot be undone}, e.g., the user cannot 'un-drop' a dataframe column. If the user executes a cell that alters the session state, a common approach to restore the previous state would be to restart the kernel and then (painstakingly) re-run past cells in the correct order.
While code versions can be saved using tools such as Git~\cite{git} or native commands (e.g., Jupyter's \%checkpoint~\cite{jupytercheckpoint}\footnote{Despite its name, \%checkpoint only stores cell code and not objects in the state.}) to simplify identifying cells to rerun for restoration, cell reruns can still be time-consuming (e.g., re-training an ML model) and/or result in incorrect restoration (e.g., random train-test splits). 
Another approach is for the user to periodically checkpoint the session state (e.g., memory dump~\cite{criu, ansel2009dmtcp} or session state serialization~\cite{dumpsession, weinman2021fork}) to storage or a managed database (e.g., KV-store~\cite{jupyterstore}).
Then, users can load an appropriate checkpoint file to restore the session state. However, performing session checkpointing and restoration using these tools is limiting: checkpointing can incur prohibitive costs (\cref{sec:experiment_cheap_checkpoint}, \cref{sec:experiment_fast_checkpoint}) and may fail on certain workloads (e.g., GPU~\cite{criucuda}), and restoration can either (1) only be (inefficiently) performed \textit{from scratch}, requiring completely loading a checkpoint file~\cite{dumpsession} and/or killing the current kernel~\cite{criu}, or (2) may be \textit{incorrect}, breaking inter-variable relations~\cite{jupyterstore}.

\paragraph{Our Goal: Generalizable, Correct, and Efficient Time-Traveling}
We propose \system, a notebook system that enables time-traveling between session states: as the user executes cells, \system tracks the session state evolution while writing per-cell incremental \textit{checkpoints} containing differing data between successive states (i.e., the \textit{state delta}) for returning to any past state via an incremental \textit{checkout} later. 
\system pursues three challenging goals---\textbf{Delta-Efficient Checkpoint:} \system aims to minimize incremental checkpointing overhead by exploiting the small per-cell deltas typical of data science workflows (\cref{sec:motivation_workload}), but also avoid high detection overhead in the face of complex access patterns and inter-variable dependencies. \textbf{Correct \& Non-intrusive Checkout:} \system aims to restore past states in the same session non-intrusively by leveraging existing objects in the kernel (that don't need updating) to minimize data loading costs, while still guaranteeing checkout accuracy as if it completely loaded a checkpoint file.
\textbf{Generalizability:} \system aims to support checkpointing/checkout for almost all notebook libraries (and/or use cases), of which there is a large variety, e.g., notebooks can perform distributed computing (e.g., Spark~\cite{zaharia2010spark}) or move data off-CPU (e.g., GPUs~\cite{pytorch}).
If \system can achieve these goals, \system will allow users to undo almost any executed cell that undesirably modifies the state as if it never occurred by quickly checking out to the pre-execution state at the cost of minimal workflow overhead.

\paragraph{Our Approach}
Our core idea for achieving our goals is to capture the state delta with low overhead, but at sufficiently high granularity using information exclusive to the application level, as follows:

First, for \textit{delta-efficient incremental checkpointing}, \system utilizes low-overhead live analysis (e.g., namespace patching) to track session state evolution at a novel \textit{Co-variable} granularity (i.e., connected components of objects).
Then, \system writes and versions Co-variables with the \textit{checkpoint graph} representing the user workflow in terms of cell executions to minimize delta storage overhead.

Second, for \textit{correct incremental checkout}, \system identifies the difference between the current and target states at the aforementioned Co-variable granularity according to the checkpoint graph.
Then, it \textit{replaces} (only) Co-variables that need updating in the state by loading data from appropriate incremental checkpoints, minimizing data load time for checkout and transparently restoring the state in the same kernel process without interruption, at \textit{sub-second latency}.


Third, \system achieves generalizability and fault-tolerance through \textit{fallback recomputation}.
If a Co-variable cannot be stored in a checkpoint (e.g., it contains an unserializable object such as a hash~\cite{hashlib}) or fails to load upon checkout, \system can efficiently reconstruct it upon checkout via finding the \textit{shortest path} combining intermediate data loading and cell re-running according to the checkpoint graph.



\paragraph{Difference from Existing Work (\cref{tbl:existing_work})} 

Our work enables efficient time-traveling for computational notebooks through significantly different techniques vs. existing work.
OS-level tools~\cite{criu, ansel2009dmtcp} can incrementally checkpoint notebook states, but fail to exploit the fine-grained deltas of data science workflows (\cref{sec:motivation_workload}), cannot incrementally restore, and fail on remote objects (e.g. Ray~\cite{moritz2018ray}, on-device data~\cite{torchtensor}).
Existing application-level tools for saving state~\cite{dill, pickle, lielasticnotebook, weinman2021fork} lack both the incremental checkpointing and restoration capabilities of our work.
Works for versioning notebook cell code~\cite{brachmann2019data, guo2012burrito, jupytercheckpoint} help identify cells to rerun for restoration but do not directly enable it.
Incremental checkpointing/PITR in DBMS~\cite{antonopoulos2019constant, mysql}/HPC~\cite{lin2020incremental, keller2019application} and time-traveling DBMS~\cite{soroush2013time, schule2019versioning} focus on robust and fast logging of table/KV-store updates, while our work focuses on delta computation for complex, interdependent objects unique to notebooks, i.e., \textbf{\textit{what to log}} (\cref{sec:motivation_workload}).
Orthogonal works include notebook systems for speeding up data exploration using non-time-travel methods (e.g., code recommendation~\cite{lee2021lux, li2023edassistant}).

\paragraph{Contributions}

According to our motivations in \cref{sec:motivation}, we implement \system (\cref{sec:system}), a notebook system with the following contributions:
\begin{itemize}
    \item \textbf{State Delta Detection.} We introduce our modeling of session state evolution at a novel \textit{Co-variable} granularity, and our correct and efficient delta detection at this granularity. (\cref{sec:checkpoint})
    
    \item \textbf{State Versioning.} We introduce our delta-based session state versioning with the \textit{Checkpoint Graph}, which enables efficient and fault-tolerant incremental checkpointing and checkout. (\cref{sec:restore})

    
    \item \textbf{Time-traveling.} 
    We show via experimental evaluation that \system's time-traveling is compatible with 146 classes from popular data science libraries and reduces checkpoint size and checkout time by up to 4.55$\times$ and 9.02$\times$, respectively. (\cref{sec:experiments})
\end{itemize}




    

    

\section{Motivation}
\label{sec:motivation}

This section describes use cases for time-traveling notebooks
(\Cref{sec:motivation_use_case}), characteristics of notebook workloads (\cref{sec:motivation_workload}), and, accordingly, our intuition for time-traveling (\cref{sec:motivation_timetravel}) and capturing state delta (\cref{sec:motivation_delta}).


\subsection{Why is Time Traveling Useful?}
\label{sec:motivation_use_case}
Time-traveling computational notebooks can enable users to efficiently undo cell executions and perform path-based exploration.

\paragraph{Undoing Cell Executions}
Data operations can be irreversible (e.g., \texttt{df=df.drop\_col('a')}), and users may want to return to the previous state if the operation outcome is undesirable~\cite{jupyterundo, brachmann2020your}.
To enable interruption-free time-traveling, we can checkpoint the state delta to storage after each operation such that the session state prior to performing the operation can be returned to via loading the appropriate deltas.
We empirically study this use case in \cref{sec:experiment_checkout_undo}.

\paragraph{Path-based exploration}
Data scientists often manually create branching, out-of-order cells, where each intended execution path consists of only a subset of notebook cells (e.g., cleaning steps for different models)~\cite{rule2018exploration, kery2018story}.
If we can efficiently persist all variations of objects in different paths as incremental deltas w.r.t. a shared state, users can efficiently switch paths for comparisons: only the (small amount of) data differing between paths need to be updated via loading deltas.
We empirically study this use case in \cref{sec:experiment_checkout_branch}.

\subsection{Characteristics of Notebook Workloads}
\label{sec:motivation_workload}
Data scientists commonly use computational notebooks for exploratory work~\cite{kery2018story}, which exhibit these key characteristics:

\paragraph{Think Time}
Notebook workflows often follow a sequential\footnote{Existing notebook systems, e.g., Jupyter, do not support concurrent cell executions.} loop of writing and running cell code, then observing output~\cite{rule2018exploration}. The notebook kernel can be inactive mid-loop---$\sim$10 seconds of \textit{think time}~\cite{eichmann2020idebench} when users decide what cells to write/run next. Notebook systems use think time for computations (e.g., pre-loading data~\cite{xin2021enhancing}).
\system can also leverage think time 
    for checkpointing.

\paragraph{Complex Access Patterns}
Python functions in notebooks can have complex access patterns (e.g., non-parameter variable accesses) that make them difficult and/or costly to analyze.
Some works use rule-based approaches to hard-code effects of common functions (e.g., \texttt{print} not modifying the namespace)~\cite{macke2020fine}. however, data scientists also often import custom functions into notebooks~\cite{rule2018exploration}; hence, an efficient approach that handles arbitrary functions is desirable.

\paragraph{Incremental Operations}
For fast iteration, users often run incremental cells each consisting of few lines of code and accessing few variables~\cite{kery2018story}: we observe this in on of our test notebooks, \textit{Sklearn} (\cref{fig:background_access_objs}), where 40/44 cells access <10\% of state data. 
For these workloads, \system's incremental checkpointing can offer
    larger benefits.

\paragraph{Data Modifications}
Notebook providers often limit session memory consumption~\cite{colabmemory, kagglememory}. Hence, while creating new data, users also modify/delete existing data to conserve memory (e.g., after they have been used to generate relevant figures): in \textit{Sklearn} (\cref{fig:background_append_update_objs}), we observe a 45:55 split between created and modified data.
\system must efficiently undo these modifications during checkout.

\begin{figure}[t]
\pgfplotsset{scaled y ticks=false}
\centering
\begin{subfigure}[b]{0.48\linewidth}
\begin{tikzpicture}

\begin{axis}[
    xtick=data,
    width=45mm,
    height=26mm,
    ymin=0,
    ymax=80,
    axis y line*=none,
    axis x line*=none,
    xtick={1,2,3,4,5, 6},
    xticklabel style   = {align=center},
    xticklabels = {1600, 800, 400, 200, 100, 50},
    ytick={0, 20, 40, 60, 80},
    yticklabels={0, 20, 40, 60, 80},
    xlabel=No. cell executions,
    xlabel style={yshift = 2.5ex},
    ylabel style={yshift=-4ex},
    xmin = 0,
    xmax = 44,
    xtick = {0, 10, 20, 30, 40},
    xticklabels = {0, 10, 20, 30, 40},
    tick label style={font=\footnotesize},
    legend style={
        at={(-0.2,1.1)},anchor=south west,column sep=2pt,
        draw=black,fill=white,
        /tikz/every even column/.append style={column sep=5pt},
        inner ysep=0.5pt,
        font=\scriptsize,
    },
    legend cell align={left},
    legend columns=4,
    label style={font=\footnotesize},
    ylabel={No. of vars},
    ymajorgrids,
]

\addplot[GreenColor, mark = x, mark size=0.75pt, opacity = 0.7, densely dashed]
table[x=cellnum,y=total, densely dashed] {sections/data/profiling_var_count.txt};
\addlegendentry{Total data in session state}
\addplot[HeuristicColor, mark = *, mark size=0.75pt, opacity = 0.7]
table[x=cellnum,y=profiled] {sections/data/profiling_var_count.txt};
\addlegendentry{Data accessed per cell}



\end{axis}
\end{tikzpicture}
\vspace{-6.5mm}
\caption{Accessed variables}
\label{fig:background_access_vars}
\end{subfigure}
\begin{subfigure}[b]{0.48\linewidth}
\begin{tikzpicture}

\begin{axis}[
    xtick=data,
    width=45mm,
    height=26mm,
    ymin=0,
    ymax=1500000000,
    axis y line*=none,
    axis x line*=none,
    xtick={1,2,3,4,5, 6},
    xticklabel style   = {align=center},
    xticklabels = {1600, 800, 400, 200, 100, 50},
    ytick={0, 500000000, 1000000000, 1500000000},
    yticklabels={0, 0.5, 1, 1.5},
    xlabel=No. cell executions,
    xlabel style={yshift = 2.5ex},
    label style={font=\footnotesize},
    ylabel style={yshift=-4ex, xshift=-0.5ex,font=\scriptsize},
    xmin = 0,
    xmax = 44,
    xtick = {0, 10, 20, 30, 40},
    xticklabels = {0, 10, 20, 30, 40},
    tick label style={font=\footnotesize},
    legend style={
        at={(-0.2,1.1)},anchor=south west,column sep=2pt,
        draw=black,fill=white,
        /tikz/every even column/.append style={column sep=5pt},
        font=\scriptsize,
    },
    legend cell align={left},
    legend columns=4,
    ylabel={Total size (GB)},
    ymajorgrids,
]

\addplot[GreenColor, mark = x, mark size=0.75pt, opacity = 0.7, densely dashed]
table[x=cellnum,y=total] {sections/data/profiling_var_size.txt};
\addplot[HeuristicColor, mark = *, mark size=0.75pt, opacity = 0.7]
table[x=cellnum,y=profiled] {sections/data/profiling_var_size.txt};



\end{axis}
\end{tikzpicture}
\label{fig:background_access_objs}
\vspace{-2.5mm}
\caption{Accessed objects}
\end{subfigure}
\hfill
\begin{subfigure}[b]{0.48\linewidth}
\begin{tikzpicture}

\begin{axis}[
    xtick=data,
    width=45mm,
    height=26mm,
    ymin=0,
    ymax=8,
    axis y line*=none,
    axis x line*=none,
    xtick={1,2,3,4,5, 6},
    xticklabel style   = {align=center},
    xticklabels = {1600, 800, 400, 200, 100, 50},
    ytick={0, 2, 4, 6, 8},
    yticklabels={0, 2, 4, 6, 8},
    xlabel=No. cell executions,
    xlabel style={yshift = 2.5ex},
    ylabel style={yshift=-4ex},
    xmin = 0,
    xmax = 44,
    xtick = {0, 10, 20, 30, 40},
    xticklabels = {0, 10, 20, 30, 40},
    tick label style={font=\footnotesize},
    legend style={
        at={(-0.2,1.1)},anchor=south west,column sep=2pt,
        draw=black,fill=white,
        /tikz/every even column/.append style={column sep=5pt},
        inner ysep=0.5pt,
        font=\scriptsize,
    },
    legend cell align={left},
    legend columns=4,
    label style={font=\footnotesize},
    ylabel={No. of vars},
    ymajorgrids,
]

\addplot[GreenColor, mark = x, mark size=0.75pt, opacity = 0.7, densely dashed]
table[x=cellnum,y=modified, densely dashed] {sections/data/profiling_var_modify_append_count.txt};
\addlegendentry{Modified data per cell}
\addplot[HeuristicColor, mark = *, mark size=0.75pt, opacity = 0.7]
table[x=cellnum,y=created] {sections/data/profiling_var_modify_append_count.txt};
\addlegendentry{Created data per cell}



\end{axis}
\end{tikzpicture}

\vspace{-2.5mm}
\caption{Modified vs. created variables}
\label{fig:background_append_update_vars}
\end{subfigure}
\begin{subfigure}[b]{0.48\linewidth}
\begin{tikzpicture}

\begin{axis}[
    xtick=data,
    width=45mm,
    height=26mm,
    ymin=0,
    ymax=1000000000,
    axis y line*=none,
    axis x line*=none,
    xtick={1,2,3,4,5, 6},
    xticklabel style   = {align=center},
    xticklabels = {1600, 800, 400, 200, 100, 50},
    ytick={0, 200000000, 400000000, 600000000, 800000000, 1000000000},
    yticklabels={0, 0.2, 0.4, 0.6, 0.8, 1},
    xlabel=No. cell executions,
    xlabel style={yshift = 2.5ex},
    label style={font=\footnotesize},
    ylabel style={yshift=-4ex, xshift=-0.5ex,font=\scriptsize},
    xmin = 0,
    xmax = 44,
    xtick = {0, 10, 20, 30, 40},
    xticklabels = {0, 10, 20, 30, 40},
    tick label style={font=\footnotesize},
    legend style={
        at={(-0.2,1.1)},anchor=south west,column sep=2pt,
        draw=black,fill=white,
        /tikz/every even column/.append style={column sep=5pt},
        font=\scriptsize,
    },
    legend cell align={left},
    legend columns=4,
    ylabel={Total size (GB)},
    ymajorgrids,
]

\addplot[GreenColor, mark = x, mark size=0.75pt, opacity = 0.7, densely dashed]
table[x=cellnum,y=modified] {sections/data/profiling_var_modify_append_size.txt};
\addplot[HeuristicColor, mark = *, mark size=0.75pt, opacity = 0.7]
table[x=cellnum,y=created] {sections/data/profiling_var_modify_append_size.txt};



\end{axis}
\end{tikzpicture}
\label{fig:background_append_update_objs}
\vspace{-2.5mm}
\caption{Modified vs. created objects}
\end{subfigure}
\caption{Pattern of a \textit{Sklearn} notebook~\cite{sklearntweet}: (Top) many cells incrementally access a small portion of the state. (Bottom) Users balance data creation and modification.
}
\label{fig:background_workload_characteristics}
\end{figure}

\subsection{Enabling Time Traveling}
\label{sec:motivation_timetravel}
We discuss the pros and cons of incremental checkpointing and checkout approaches for enabling time-traveling to a previous state.

\paragraph{OS-level Memory Snapshots} Tools such as CRIU~\cite{criu} snapshot notebook process memory to save session state data. \textbf{(Incremental Checkpointing)} Subsequent snapshots can be made incrementally w.r.t. prior snapshots storing only dirty memory pages. However, memory-page granularity is too coarse for notebook workloads as Python data structures (e.g., lists) can be fragmented, on which operations (e.g., in-place list sorting) can cause multiple dirty pages and high checkpoint costs (\cref{sec:experiment_cheap_checkpoint}).
\textbf{(Complete Checkout)} Memory snapshots must be entirely loaded for state restoration and require killing the existing notebook process (to avoid PID conflicts) before restoration, which is not seamless and incurs high data loading costs (\cref{sec:experiment_checkout}). OS-level snapshotting is also limited to single processes, hence fail on notebooks utilizing multiple/remote processors (\cref{sec:experiment_cheap_checkpoint}).
\paragraph{Application-level Time-Traveling (Ours)}
We use (1) application-level information to detect state deltas at a finer granularity versus memory snapshots and (2) existing data in the kernel for incremental checkout to address the drawbacks of OS-level snapshotting: 
\textbf{(Incremental and Generalized Checkpointing)} We track state deltas at a 
more logical
\textit{Co-variable granularity} (described in \cref{sec:motivation_delta}) by tracking in-notebook references for incremental checkpointing. For generalizability, we use objects' \textit{reductions} as storage instructions to checkpoint multiprocessing and off-CPU workloads (\cref{sec:implementation_details}).
\textbf{(Incremental Checkout)} As we can directly access the notebook kernel at the application level, we can incrementally checkout by computing the difference between the current and target states (e.g., via versioning~\cite{bernstein1983multiversion}) and updating only differing kernel data (\cref{sec:restore_fast}).

\subsection{Tracking State Delta for Time Traveling}
\begin{figure}[t]

\centering
\begin{tikzpicture}[>={LaTeX[width=1mm,length=1mm]},->]

\node(notebook) [draw=black, anchor=north west, minimum width = 31mm, minimum height=19.5mm,densely dashdotdotted]
at (0,0) {};
\node(nbtitle) [anchor=south west, minimum height=3mm, inner sep = 0mm]
at ($(notebook.north west) + (0, 0)$) {\small \textbf{Notebook}};

\node(cell3) [draw=black, anchor=north, minimum width = 26mm, minimum height=3.5mm]
at ($(notebook.north) + (0.15, -0.1)$) {};
\node(celltext3) [anchor=north east, minimum height=3mm, inner sep = 0.2mm]
at ($(cell3.north west) + (0, 0)$) {\footnotesize [1]};
\node(codetxt3) [anchor=west, minimum height=3mm, inner sep = 0.2mm,align=left]
at ($(cell3.west) + (0.05, 0)$) {\footnotesize\texttt{df = read\_csv(\textcolor{purple}{'...'})}};

\node(cell) [draw=black, anchor=north, minimum width = 26mm, minimum height=6mm]
at ($(cell3.south) + (0, -0.1)$) {};
\node(celltext3) [anchor=north east, minimum height=3mm, inner sep = 0.2mm]
at ($(cell.north west) + (0, 0)$) {\footnotesize [2]};
\node(codetxt) [anchor=west, minimum height=6mm, inner sep = 0.2mm,align=left]
at ($(cell.west) + (0.05, 0)$) {\footnotesize\texttt{ser=pd.Series(}\\[-0.35em]\,\,\footnotesize\texttt{[\textcolor{purple}{'a'}, \textcolor{purple}{'b'}, \textcolor{purple}{'c'}])}};
\node(cell2) [draw=black, anchor=north, minimum width = 26mm, minimum height=6mm]
at ($(cell.south) + (0, -0.1)$) {};
\node(celltext3) [anchor=north east, minimum height=3mm, inner sep = 0.2mm]
at ($(cell2.north west) + (0, 0)$) {\footnotesize [3]};
\node(codetxt2) [anchor=west, minimum height=3mm, inner sep = 0.2mm,align=left]
at ($(cell2.west) + (0.05, 0)$) {\footnotesize\texttt{obj = MyObj(foo=}\\[-0.35em]\footnotesize \texttt{ser[1],bar=\textcolor{purple}{'d'})}};

\node(magictxt) [
    anchor=west, minimum height=19.5mm, minimum width = 42mm, inner sep = 2mm, align=left, font=\ttfamily\footnotesize,
    draw=black, rounded corners=1mm,
]
at ($(notebook.east) + (1.1, 0.0)$) {};
\node(magictitle) [anchor=south west, minimum height=3mm, inner sep = 0mm]
at ($(magictxt.north west) + (0, 0)$) {\small \textbf{Namespace}};
\node(magictxt2) [
    anchor=north west, align=left, font=\ttfamily\footnotesize,
]
at ($(magictxt.north west) + (0.1, 0)$) {
    \textcolor{RandomColor}{\# Identify Co-variables in}\\[-0.2em]\textcolor{RandomColor}{\# Namespace}
};

    \node(objx) [draw=black, fill=vintageblue, fill opacity = 0.6,text opacity=1, thick, anchor=west, minimum width = 5mm, minimum height=5mm, circle, inner sep = 0.4mm]
 at ($(magictxt.south west) + (0.2, 0.35)$) {\scriptsize{data}};

 \node(objy) [draw=black, fill=vintageblue, fill opacity = 0.6,text opacity=1, thick, anchor=west, minimum width = 5mm, minimum height=5mm, circle, inner sep = 0.4mm]
 at ($(objx.east) + (0.3, 0)$) {\scriptsize{idx}};

 \node(objx2) [draw=black, fill=vintageblue, fill opacity = 0.6,text opacity=1, thick, anchor=south, minimum width = 5mm, minimum height=5mm, circle, inner sep = 0.4mm]
 at ($(objx.north) + (0.4, 0.2)$) {\small{df}};

  \node(obja) [draw=black, fill=vintagered, fill opacity = 0.6,text opacity=1, thick, anchor=west, minimum width = 3mm, minimum height=3mm, circle, inner sep = 0mm]
 at ($(objy.east) + (0.25, 0)$) {\scriptsize{'a'}};
   \node(objb) [draw=black, fill=vintagered, fill opacity = 0.6,text opacity=1, thick, anchor=west, minimum width = 3mm, minimum height=3mm, circle, inner sep = 0mm]
 at ($(obja.east) + (0.3, 0)$) {\scriptsize{'b'}};
   \node(objc) [draw=black, fill=vintagered, fill opacity = 0.6,text opacity=1, thick, anchor=west, minimum width = 3mm, minimum height=3mm, circle, inner sep = 0mm]
 at ($(objb.east) + (0.3, 0)$) {\scriptsize{'c'}};
   \node(objd) [draw=black, fill=vintagered, fill opacity = 0.6,text opacity=1, thick, anchor=west, minimum width = 3mm, minimum height=3mm, circle, inner sep = 0mm]
 at ($(objc.east) + (0.3, 0)$) {\scriptsize{'d'}};
    \node(objls) [draw=black, fill=vintagered, fill opacity = 0.6,text opacity=1, thick, anchor=south, minimum width = 5mm, minimum height=5mm, circle, inner sep = 0.4mm]
 at ($(objb.north) + (0, 0.2)$) {\scriptsize{ser}};
     \node(objobj) [draw=black, fill=vintagered, fill opacity = 0.6,text opacity=1, thick, anchor=south, minimum width = 5mm, minimum height=5mm, circle, inner sep = 0.4mm]
 at ($(objc.north) + (0, 0.2)$) {\scriptsize{obj}};

\draw[->, thick] 
($(objls.south)$) --
($(obja.north)$); 
\draw[->, thick] 
($(objls.south)$) --
($(objb.north)$); 
\draw[->, thick] 
($(objls.south)$) --
($(objc.north)$); 
\draw[->, thick] 
($(objobj.south)$) --
($(objb.north)$); 
\draw[->, thick] 
($(objobj.south)$) --
($(objd.north)$); 
\draw[->, thick] 
($(objx2.south)$) --
($(objx.north)$); 
\draw[->, thick] 
($(objx2.south)$) --
($(objy.north)$); 

\draw[line width=1.0mm,-latex]  
    ($(notebook.east) + (0.2, 0.2)$) -- ($(notebook.east) + (1.0, 0.2)$);

\end{tikzpicture}

\caption{Co-variables are connected components of objects. We can treat them as independent data tables.}
\label{fig:background_variable_blob}
\end{figure}
\label{sec:motivation_delta}
We discuss pros and cons of methods of tracking the state delta at the application level for incremental checkpointing and checkout.

\paragraph{Provenance-based Tracking}
Some existing notebook systems~\cite{macke2020fine, koop2017dataflow, lielasticnotebook} track state deltas via provenance-based code analysis (e.g., via ASTs~\cite{ast}) at variable-level granularity.
Pure static analysis requires conservativeness on identifying changed data w.r.t. control flows (e.g., \texttt{if(x<1))} and external function calls (\cref{sec:motivation_workload}), causing false positives (e.g., assuming an untaken branch as taken) and large deltas; hence, these systems augment static analysis with varying levels of live instrumentation at cell runtime (e.g., resolving \texttt{x}'s value when evaluating \texttt{if(x<1)})~\cite{macke2020fine}, which can result in high overhead (e.g., repeated resolutions in loops, \cref{sec:experiment_fast_capture}).

\paragraph{Co-variable Granularity Live Tracking (Ours)}

To avoid potential inefficiencies of provenance tracking's runtime resolutions, we propose performing \textit{live object comparison} (i.e., comparing data pre/post-execution) only between cell executions to track per-cell updates.
Our intuition is that while doing so at variable-level granularity (like existing provenance trackers) by comparing all state objects can be costly (\cref{sec:experiment_fast_capture}), it is also unnecessary as storing/loading individual variables (e.g., with variable-level KV-stores~\cite{pythonshove, redisshelve}) for checkpoint/checkout risks breaking shared references~\cite{lielasticnotebook}.
We hence track updates at a coarser \textit{Co-variable} level---connected components of objects (w.r.t. pointer references): we can reason from access patterns which Co-variables were possibly/surely not updated by each cell to limit object comparisons, and correctly store/load Co-variables as if they are independent data tables. \cref{fig:background_variable_blob} depicts our idea: \texttt{\{ser,obj\}} is a Co-variable (red) as objects reachable from \texttt{ser} and \texttt{obj} overlap (\texttt{\&ser[1]=\&obj.foo}). \texttt{\{df\}} is another Co-variable (blue), and users \textit{cannot} reach objects under \texttt{df} from objects under \texttt{ser}/\texttt{obj} via references. Notably, Co-variables are the \textit{minimum granularity} for saving/loading state data without risking breaking shared references.
\cref{sec:checkpoint} formally describes Co-variables and how we correctly and efficiently capture state delta at this granularity.

\paragraph{Motivating Example (\cref{fig:background_example})}
A data analyst performs text mining by loading the corpus (Cell 1), defining category lists (Cell 2), and sorting texts by sentiment into the lists (Cell 3). They checkpoint the state after each cell execution.
\begin{figure}[t]

\centering
\begin{tikzpicture}[>={LaTeX[width=1mm,length=1mm]},->]

\node(notebook) [draw=black, anchor=north west, minimum width = 39mm, minimum height=29.5mm,densely dashdotdotted]
at (0,0) {};
\node(notebooktxt) [anchor=south west, minimum height=3mm, inner sep = 0mm]
at ($(notebook.north west) + (0, 0)$) {\small \textbf{Notebook}};
\node(cell) [draw=black, anchor=north, minimum width = 34mm, minimum height=3.5mm]
at ($(notebook.north) + (0.15, -0.1)$) {};
\node(celltext) [anchor=north east, minimum height=3mm, inner sep = 0.2mm]
at ($(cell.north west) + (0, 0)$) {\footnotesize [1]};
\node(codetxt) [anchor=west, minimum height=3mm, inner sep = 0.6mm]
at ($(cell.west) + (0.05,0)$) {\footnotesize \texttt{corpus=read\_csv('...')}};
\node(cell2) [draw=black, anchor=north, minimum width = 34mm, minimum height=7mm]
at ($(cell.south) + (0, -0.1)$) {};
\node(cell2text) [anchor=north east, minimum height=3mm, inner sep = 0.2mm]
at ($(cell2.north west) + (0, 0)$) {\footnotesize [2]};
\node(codetxt2) [anchor=west, minimum height=3mm, inner sep = 0.6mm, align=left]
at ($(cell2.west) + (0.05, 0)$) {\footnotesize \texttt{sad\_ls = []}\\[-0.35em]\footnotesize \texttt{happy\_ls = []}\\[-0.8em]\footnotesize \texttt{...}};
\node(cell3) [draw=black, anchor=north, minimum width = 34mm, minimum height=10mm]
at ($(cell2.south) + (0, -0.1)$) {};
\node(celltext3) [anchor=north east, minimum height=3mm, inner sep = 0.2mm]
at ($(cell3.north west) + (0, 0)$) {\footnotesize [3]};
\node(codetxt3) [anchor=west, minimum height=3mm, inner sep = 0.6mm, align=left]
at ($(cell3.west) + (0.05, 0)$) {\footnotesize \texttt{for row in corpus:}\\[-0.35em]\footnotesize \texttt{\,\,if row['mood']=='sad':}\\[-0.35em]\footnotesize \texttt{\,\,\,\,\textcolor{purple}{sad\_ls}.append(row['txt'])}\\[-0.8em]\footnotesize \texttt{...}};

\node(cell4) [draw=black, anchor=north, minimum width = 34mm, minimum height=3.5mm]
at ($(cell3.south) + (0, -0.1)$) {};
\node(celltext4) [anchor=north east, minimum height=3mm, inner sep = 0.2mm]
at ($(cell4.north west) + (0, 0)$) {\footnotesize [4]};
\node(codetxt4) [anchor=west, minimum height=3mm, inner sep = 0.6mm, align=left]
at ($(cell4.west) + (0.05, 0)$) {\footnotesize \texttt{\textcolor{vintageblue}{sad\_ls} = [re.sub('r$\setminus$W')...}};

\node(nstxt) [anchor=west, inner sep = 0mm]
at ($(notebooktxt.east) + (2.7, 0)$) {\small \textbf{Namespace}};

\node(corpus1) [draw=black, fill = Lightgrey, anchor=north west, minimum height=3.5mm, minimum width = 19mm,inner sep = 0.6mm]
at ($(cell.north east) + (0.2, 0.0)$) {\footnotesize \texttt{corpus}};
\node(corpus2) [draw=black, fill = Lightgrey, anchor=north west, minimum height=3.5mm, minimum width = 19mm,inner sep = 0.6mm]
at ($(cell2.north east) + (0.2, 0.0)$) {\footnotesize \texttt{corpus}};
\node(corpus3) [draw=black, fill = Lightgrey, anchor=north west, minimum height=3.5mm, minimum width = 19mm,inner sep = 0.6mm]
at ($(cell3.north east) + (0.2, -0.15)$) {\footnotesize \texttt{corpus}};
\node(corpus4) [draw=black, fill = Lightgrey, anchor=north west, minimum height=3.5mm, minimum width = 19mm,inner sep = 0.6mm]
at ($(cell4.north east) + (0.2, 0.0)$) {\footnotesize \texttt{corpus}};

\node(sadls2) [draw=black, fill = Lightgrey, anchor=west, minimum height=3.5mm, minimum width = 2mm,inner sep = 0.6mm]
at ($(corpus2.east) + (0.1, 0.0)$) {};
\node(happyls2) [draw=black, fill = Lightgrey, anchor=west, minimum height=3.5mm, minimum width = 2mm,inner sep = 0.6mm]
at ($(sadls2.east) + (0.1, 0.0)$) {};
\node[] (nsdots2) at ($(happyls2.east) + (0.2, 0)$) {\large{\bf ...}};

\node(sadls3) [draw=black, fill = vintagered, anchor=west, minimum height=3.5mm, minimum width = 2mm,inner sep = 0.6mm]
at ($(corpus3.east) + (0.1, 0.0)$) {\footnotesize \texttt{sad\_ls}};
\node(happyls3) [draw=black, fill = Lightgrey, anchor=west, minimum height=3.5mm, minimum width = 2mm,inner sep = 0.6mm]
at ($(sadls3.east) + (0.1, 0.0)$) {\footnotesize \texttt{happy\_ls}};
\node[] (nsdots3) at ($(happyls3.east) + (0.2, 0)$) {\large{\bf ...}};

\node(sadls4) [draw=black, fill=vintageblue, fill opacity = 0.6,text opacity=1, anchor=west, minimum height=3.5mm, minimum width = 2mm,inner sep = 0.6mm]
at ($(corpus4.east) + (0.1, 0.0)$) {\footnotesize \texttt{sad\_ls}};
\node(happyls4) [draw=black, fill = Lightgrey, anchor=west, minimum height=3.5mm, minimum width = 2mm,inner sep = 0.6mm]
at ($(sadls4.east) + (0.1, 0.0)$) {\footnotesize \texttt{happy\_ls}};
\node[] (nsdots4) at ($(happyls4.east) + (0.2, 0)$) {\large{\bf ...}};

\node(checkouttxt) [anchor=east, minimum height=6mm, inner sep = 0.2mm,align=right]
at ($(sadls4.north) + (0.1, 0.3)$) {\footnotesize \textbf{(2)} Incremental checkout\\[-0.35em]\footnotesize by only updating sad\_ls};

\draw[->, thick] 
($(sadls4.north) + (0.2, 0.1)$) --
($(sadls3.south) + (0.2, -0.1)$); 

\draw[-, thick] 
($(sadls3.north) + (0.2, 0.05)$) --
($(sadls3.north) + (0.4, 0.15)$); 

\draw[-, thick] 
($(sadls3.north) + (1.9, 0.15)$) --
($(sadls3.north) + (0, 0.15)$); 

\draw[-, thick] 
($(sadls3.north) + (1.9, 0.3)$) --
($(sadls3.north) + (0, 0.3)$);

\node(checkouttxt) [draw = black, anchor=south, inner sep = 0.2mm,align=center]
at ($(sadls3.north) + (1.3, 0.6)$) {\footnotesize \texttt{b'sad\_ls'}};

\draw[->, thick] 
($(checkouttxt.south) + (0, -0.2)$) --
($(checkouttxt.south) + (0, -0.05)$); 

\node(checkouttxt2) [anchor=east, inner sep = 0.2mm,align=center]
at ($(sadls3.north) + (-0.05, 0.3)$) {\footnotesize \textbf{(1)} efficient Co-variable \\[-0.35em]\footnotesize level incremental ckpt.};

\node(fragment2) [draw=black, fill = vintagered, anchor=west, minimum height=1.5mm, minimum width = 2mm,inner sep = 0mm]
at ($(sadls3.north) + (0.3, 0.225)$) {};
\node(fragment3) [draw=black, fill = vintagered, anchor=west, minimum height=1.5mm, minimum width = 2mm,inner sep = 0mm]
at ($(sadls3.north) + (0.9, 0.225)$) {};
\node(fragment4) [draw=black, fill = vintagered, anchor=west, minimum height=1.5mm, minimum width = 2mm,inner sep = 0mm]
at ($(sadls3.north) + (1.6, 0.225)$) {};
\end{tikzpicture}
\caption{Co-variable granularity deltas allows us to create size-efficient incremental checkpoints (vs. memory-page level deltas), and incrementally checkout to previous states.}
\label{fig:background_example}
\end{figure}
\textbf{Incremental Checkpointing:} The analyst tests a mapping to clean the contained text in the lists on \texttt{sad\_ls} (Cell 4, blue).
Due to its interleaved construction (with other lists), \texttt{sad\_ls} is fragmented; incrementally checkpointing at memory page granularity for Cell 4 (w.r.t. Cell 3) copies all pages overlapping with \texttt{sad\_ls}.
However, a Co-variable granularity incremental checkpoint stores only (the bytestring of) \texttt{sad\_ls}. \textbf{Incremental Checkout:}
The analyst undoes Cell 4's mapping due to poor results. Returning to Cell 3's state by (completely) loading a memory snapshot is slow as it also reloads the corpus.
However, noting that Cell 3's and 4's states differ only by \texttt{sad\_ls}, we can only load Cell 3's \texttt{sad\_ls} (red) to replace Cell 4's \texttt{sad\_ls} (blue) to incrementally checkout without touching the rest of the state.

\section{System Overview}
\label{sec:system}

This section presents \system components (\cref{sec:system_components}) and workflow (\cref{sec:system_workflow}).


\subsection{\systemnosf Components}
\label{sec:system_components}


\system (\cref{fig:system_overview}) interacts with notebook sessions via non-intrusive hooks, which allow \system to transparently (1) monitor the namespace to track session state evolution, (2) write state data to storage for checkpointing, and (3) alter the state on requested checkouts.


\paragraph{Patched Namespace} 
On session start, \system \textit{patches} the namespace to monitor accesses to its contents during cell executions (\cref{sec:checkpoint_fast}). It tracks user-referenced variable names to identify candidate Co-variables to check for updates in, which are passed to the Delta Detector to compute the Co-variable granularity state delta.
\begin{figure}[t]
\usetikzlibrary{calc}

\begin{subfigure}{\columnwidth}
\centering

\tikzset{
mylabel/.style={
    font=\footnotesize\sffamily\bfseries,
    align=center,
},
mylabel2/.style={
    font=\footnotesize\sffamily,
    align=center,
},
mycomponent/.style={
    semithick, rounded corners=0.5mm,
}
}

\begin{tikzpicture}[>={LaTeX[width=1mm,length=1mm]},->]
\node[draw=black,thick, fill = Lightgrey, minimum width=24mm, minimum height=25mm] (B) at (0, 0) {};
\node[anchor=south,font=\small\bfseries] at (B.north) 
    {Notebook Interface};

\node(cell2) [draw=black, anchor=north east, minimum width = 19mm, minimum height=6mm, fill = white]
at ($(B.north east) + (-0.1, -0.1)$) {};
\node(codetxt2) [anchor=west, minimum height=3mm, inner sep = 0.6mm, align=left]
at ($(cell2.west) + (0.05, 0)$) {\footnotesize \texttt{import pandas}\\[-0.35em]\footnotesize \texttt{df=pd.read\_csv}};

\node(cell3) [draw=black, anchor=north east, minimum width = 19mm, minimum height=3.5mm, fill = white]
at ($(cell2.south east) + (0, -0.1)$) {};
\node(codetxt3) [anchor=west, minimum height=3mm, inner sep = 0.6mm, align=left]
at ($(cell3.west) + (0.05, 0)$) {\footnotesize \texttt{df.head}};

\node(cell4) [anchor=north west, minimum width = 14mm, minimum height=7mm, fill = white]
at ($(cell3.south west) + (0, -0.05)$) {};

\node[anchor=east,font=\footnotesize] at ($(cell2.north west) + (0.1, -0.15)$) 
    {[1]};
\node[anchor=east,font=\footnotesize] at ($(cell3.north west) + (0.1, -0.15)$) 
    {[2]};

\node[anchor=west,font=\footnotesize] at ($(cell3.south west) + (0, -0.15)$) 
    {col1};
\node[anchor=west,font=\footnotesize] at ($(cell3.south west) + (0.55, -0.15)$) 
    {col2};
\node[anchor=west,font=\footnotesize] at ($(cell3.south west) + (0, -0.4)$) 
    {1};
\node[anchor=west,font=\footnotesize] at ($(cell3.south west) + (0.55, -0.4)$) 
    {hello};
\node(commandbox) [draw=black, fill=white, anchor=south, minimum width = 24mm, minimum height=8mm, thick]
at ($(B.south) + (0, 0)$)  {};

\node(commandpalette) [anchor=north, align = center, minimum height=3mm, inner sep = 0.5mm, mylabel]
at ($(commandbox.north) + (0, -0.05)$) {\scriptsize \system command palette};

\node(cmdline) [draw=black, fill=white, anchor=south, minimum width = 22mm, minimum height=3.5mm, thick]
at ($(commandbox.south) + (0, 0.1)$)  {};

\node(cmd) [anchor=west, minimum height=3mm, inner sep = 0.6mm, align=left]
at ($(cmdline.west) + (0.05, 0)$) {\footnotesize \texttt{\textbf{init/log...}}};

\draw[-] 
($(cmd.north) + (0.8, -0.1)$) --
($(cmd.south) + (0.8, 0.1)$); 
\draw[-] 
($(cmd.north) + (0.75, -0.1)$) --
($(cmd.north) + (0.85, -0.1)$); 
\draw[-] 
($(cmd.south) + (0.75, 0.1)$) --
($(cmd.south) + (0.85, 0.1)$);

\draw[-] 
($(cell3.south west) + (0, -0.27)$) --
($(cell3.south west) + (1.4, -0.27)$); 

\node(shell) [draw=black, anchor=west, minimum width = 10mm, minimum height=10mm, thick]
at ($(B.east) + (0.3, 0)$)  {};
\node(shelltxt) [anchor=center, align = center, minimum height=3mm, inner sep = 0.5mm, font=\bfseries\footnotesize]
at ($(shell.center) + (0, 0)$) {Jupyter \\ Kernel};

\node(elastic) [minimum height=25mm, minimum width=42mm,
  draw=black, anchor=west, thick, rounded corners=1mm]
at ($(shell.east) + (0.3, 0)$)  {};
\node(elastictxt) [anchor=south, minimum height=3mm, inner sep = 0mm]
at ($(elastic.north) + (0, 0.05)$) {\textbf{\system}};

\draw[<->, thick] 
($(B.east)$) --
($(shell.west)$);
\draw[<->, thick] 
($(shell.east)$) --
($(elastic.west)$);
\draw[<->, thick,dashed] 
($(cmdline.east) + (0.1, 0)$) --
($(cmdline.east) + (1.75, 0)$);

\node(intercepter) [draw=black, fill = white, anchor=north east, minimum width = 18mm, minimum height=9.7mm, mycomponent]
at ($(elastic.north east) + (-0.2, -0.2)$)  {};

\node(interceptertxt) [anchor=north, align = center, minimum height=3mm, inner sep = 0.5mm, mylabel]
at ($(intercepter.north) + (0, 0)$) {Delta \\[-0.2em] Detector (\cref{sec:checkpoint})};

\node(idgraph) [draw=black, fill = white, anchor=south, minimum width = 16mm, minimum height=2.5mm, align=center, inner sep = 0.5mm, mylabel2]
at ($(intercepter.south) + (0, 0.1)$) {\vargraphs};

\node(restorer) [draw=black, fill = white, anchor=south west, minimum width = 18mm, minimum height=6mm, mycomponent]
at ($(elastic.south west) + (0.2, 0.2)$)  {};

\node(restorertxt) [anchor=north, align = center, minimum height=3mm, inner sep = 0.5mm, mylabel]
at ($(restorer.north) + (0, -0.05)$) {Data \\[-0.2em] Restorer (\cref{sec:restore_robust})};

\node(stateloader) [draw=black, fill = white, anchor=south, minimum width = 18mm, minimum height=6mm, mycomponent]
at ($(restorer.north) + (0, 0.125)$)  {};

\node(stateloadertxt) [anchor=north, align = center, minimum height=3mm, inner sep = 0.5mm, mylabel]
at ($(stateloader.north) + (0, -0.05)$) {State \\[-0.2em] Loader (\cref{sec:restore_fast})};



\node(optimizer) [draw=black, fill = white, anchor=north west, minimum width = 18mm, minimum height=6mm, mycomponent]
at ($(elastic.north west) + (0.2, -0.2)$)  {};

\node(optimizertxt) [anchor=north, align = center, minimum height=3mm, inner sep = 0.5mm, mylabel]
at ($(optimizer.north) + (0, -0.05)$) {Patched Name-\\[-0.2em] space (\cref{sec:checkpoint_fast})};

\node(replicator) [draw=black, fill = white, anchor=south east, minimum width = 18mm, minimum height=9.7mm, mycomponent]
at ($(elastic.south east) + (-0.2, 0.2)$)  {};

\node(replicatortxt) [anchor=north, align = center, minimum height=3mm, inner sep = 0.5mm, mylabel]
at ($(replicator.north) + (0, 0)$) {Checkpoint \\[-0.2em] Graph (\cref{sec:checkpoint_manage})};

\node(writer) [draw=black, fill = white, anchor=south, minimum width = 16mm, minimum height=3mm, align=center, inner sep = 0.5mm, mylabel2]
at ($(replicator.south) + (0, 0.1)$) {Data Writer};

\end{tikzpicture}
\end{subfigure}
\caption{\systemnosf architecture. It utilizes a hook to observe session state deltas and transparently write/replace data in the kernel namespace for incremental checkpoint/checkout.}
\label{fig:system_overview}
\end{figure}
\paragraph{Delta Detector}
The Delta Detector computes the state delta based on the candidates identified from the Patched Namespace (i.e., which of the candidate Co-variables were actually updated by the cell execution). We discuss the \system's delta detection in (\cref{sec:checkpoint}).

\paragraph{Checkpoint Graph}

The Checkpoint Graph is a tree-like structure analogous to Git's commit graph~\cite{commitgraph}, in which \system writes, stores, and versions incremental checkpoints consisting of the updated Co-variables (i.e., the state delta) of each cell execution (\cref{sec:checkpoint_manage}).
The incremental checkpoints stored in the Checkpoint Graph are used by the State Loader to perform incremental checkout.

\paragraph{State Loader}
The State Loader restores to a session state upon requested checkout.
It first identifies the difference between the current (i.e., existing items in the namespace) and target states via the Checkpoint Graph, then loads necessary data from the Checkpoint Graph to replace Co-variables that need updating (\cref{sec:restore_fast}).

\paragraph{Data Restorer}
The Data Restorer is a mechanism that utilizes fallback recomputation to restore missing data for checkout (e.g., \system failed to serialize the data during prior checkpointing).
It reconstructs missing data by combining loading dependent data and cell re-runs according to the Checkpoint Graph. (\cref{sec:restore_robust})

\subsection{\systemnosf Workflow}
\label{sec:system_workflow}

This section covers \system's operations during a notebook workflow. Users interact with \system with the in-Jupyter Command Palette (\cref{fig:system_overview}), such as to \textit{attach} it to a new notebook session;\footnote{A depiction of \system's interface can be found in our prior work's demo paper~\cite{li2024demonstration}.}
\system then monitors the namespace for state deltas, checkpoint after each cell execution, and perform on-demand checkout to previous states.

\paragraph{Attaching \systemnosf to a Notebook Session}
When initializing a notebook session, \textbf{\texttt{init}} attaches \system to the kernel, which will patch the namespace and initialize the Checkpoint Graph on storage.

\paragraph{Incremental Checkpointing}
After each cell execution, the Delta Detector uses the Patched Namespace to identify updated Co-variables and stores them in a new incremental checkpoint/node with a unique \texttt{ckpt\_id} on the Checkpoint Graph.
The user may view the stored graph, checkpoints, and their IDs with \textbf{\texttt{log}}.

\paragraph{Incremental Checkout}
\system will restore a previous session state with \textbf{\texttt{checkout}} \texttt{ckpt\_id}.
The State Restorer identifies differing Co-variables between the current and target states according to the Checkpoint Graph, then accordingly loads only the necessary data for restoration.
If necessary, the Data Restorer reconstructs data that is missing or failed to load via fallback recomputation.

\section{Accurate and Fast Delta Detection}
\label{sec:checkpoint}

This section describes how \system correctly detects Co-variable granularity state deltas necessary for incremental checkpointing and checkout (\cref{sec:motivation_delta}).
We formally describe Co-variables in \cref{sec:checkpoint_blob} and how we detect Co-variable updates correctly (\cref{sec:checkpoint_detect}) efficiently (\cref{sec:checkpoint_fast}).

\subsection{Co-variables}
\label{sec:checkpoint_blob}
This section introduces \system's definition of the Co-variable.
\subfile{plots/fig_checkpoint_variable_blob_update}
\paragraph{Preliminary}
In Python and the Jupyter Notebook ecosystem, variables and objects are 2 distinct concepts: A \textbf{variable} is a named entity from which various \textbf{objects} are \textit{reachable}---for an example list \texttt{ls=[1,2,3]}, the list name (\texttt{ls}) is a variable and each list element (1, 2, 3) is an object. We define reachability reference-wise, i.e., object \texttt{y} is reachable from variable \texttt{x} if \texttt{y} can be accessed from \texttt{x} through a chain of references. Some common reachability patterns include \textit{subscripting} (e.g., \texttt{y=x[0]}) and \textit{class member} (e.g., \texttt{y=x.attr}). Given this distinction between variables and objects and reachability definition, we now define Co-variables as follows:


\begin{definition}
    A \textbf{Co-variable} is a set of variable names $\mathcal{X} = \{x_1,...,x_i\}$ from which the reachable objects form a \textbf{maximally connected component}. That is, for any variable $y$ not in the set, the objects reachable from $x_1,...,x_i$ are not reachable from $y$.
\end{definition}

A Co-variable can consist of one name (e.g., a primitive, \texttt{x=1}) or multiple names from which same objects can be reached (i.e., shared references). For example, in \cref{fig:checkpoint_variable_update}, the string object \texttt{'b'} is reachable from both Pandas Series \texttt{ser} and object \texttt{obj} via subscript and class member respectively, hence \texttt{\{ser,obj\}} is a Co-variable.
Co-variables are self-contained by definition, i.e., there are no inter-Co-variable references. They can be modified by cell executions:

\begin{definition}
    A Co-variable $\mathcal{X} = \{x_1,...,x_i\}$ is \textbf{modified} by a cell execution if the graph structure of the connected component of objects reachable from $x_1,...,x_i$ is modified, counting both node (i.e., object) and edge (i.e., reference) additions and deletions.
\end{definition}
For example, the Co-variable \texttt{\{ser,obj\}} in \cref{fig:checkpoint_variable_update} is modified node-wise with an in-place update ``\texttt{ser.replace''} \textbf{(bottom)}, and edge-wise with a re-assignment ``\texttt{obj.foo=ser[2]}'' \textbf{(bottom-right)}.
Co-variables can be created and deleted via \textit{split and merge} \textbf{(right)}: \texttt{\{obj,ser\}} is deleted via a split as \texttt{obj} and \texttt{ser} longer share references, and \texttt{\{obj,st\}} is created via a merge.
We collectively refer to Co-variable modifications, creations, and deletions as \textit{updates}; the Co-variables updated by a cell execution form its \textit{state delta}.

\subsection{Accurate State Delta Detection}
\label{sec:checkpoint_detect}

This section describes how \system accurately detects Co-variable membership (i.e., which variables form a Co-variable) and updates.

\paragraph{\vargraphs}
\system detects Co-variable membership and updates with \vargraphs---a graph structure constructed on each variable that captures its reachable objects in the namespace.
\cref{fig:checkpoint_id_graph} shows an example: each node in a variable's \vargraph corresponds to a reachable object, containing the (1) type, (2) memory address, and one of (3) pointers to other reachable objects (i.e., children) for non-primitives, or (4) value for primitives. For example, the node for \texttt{list} contains 3 child pointers to the 3 nodes for strings \texttt{'a'}, \texttt{'b'}, and \texttt{'c'}, and the node for string \texttt{'b'} holds its value \texttt{'b'}.
\footnote{The \vargraph is inspired by ElasticNotebook's \textit{ID graph}~\cite{lielasticnotebook} which captures reachable objects' memory addresses; \vargraphs uniquely contain datatypes and primitive values for additional robustness (e.g., detecting a different primitive in the same address).}
\paragraph{Detecting Co-variable membership}
Co-variable membership is determined by intersecting \vargraphs. For example, in \cref{fig:checkpoint_id_graph}, \texttt{ser} and \texttt{obj} form a Co-variable as the node \texttt{'b'} is in both graphs (red).
\begin{figure}[t]
\begin{subfigure}[b]{\linewidth}
\centering
\hspace{-3mm}
\begin{tikzpicture}[>={LaTeX[width=1mm,length=1mm]},->]

\tikzset{
mylabel/.style={
    font=\footnotesize\sffamily\bfseries,
    align=center,
},
mylabel2/.style={
    font=\footnotesize\sffamily,
    align=center,
},
mycomponent/.style={
    semithick, rounded corners=0.5mm, inner ysep=0.4mm
}
}

  \node(obja) [draw=black, fill = white, anchor=north west, align=left, minimum width = 12mm, minimum height=7.5mm, mycomponent]
 at (0, 0) {\footnotesize Type:\texttt{str}\\[-0.4em]\footnotesize Addr:\texttt{0xaaa0}\\[-0.4em]\footnotesize Val:\texttt{'a'}};
   \node(objb) [draw=BlueColor, fill = white, anchor=west, align=left, minimum width = 12mm, minimum height=7.5mm, mycomponent]
 at ($(obja.east) + (0.4, 0)$) {\footnotesize \textcolor{BlueColor}{Type:\texttt{str}}\\[-0.4em]\footnotesize \textcolor{BlueColor}{Addr:\texttt{0xbbb0}}\\[-0.4em]\footnotesize \textcolor{BlueColor}{Val:\texttt{'b'}}};
   \node(objc) [draw=black, fill = white, anchor=west, align=left, minimum width = 12mm, minimum height=7.5mm, mycomponent]
 at ($(objb.east) + (0.4, 0)$) {\footnotesize Type:\texttt{str}\\[-0.4em]\footnotesize Addr:\texttt{0xccc0}\\[-0.4em]\footnotesize Val:\texttt{'c'}};
   \node(objd) [draw=black, fill = white, anchor=west, align=left, minimum width = 12mm, minimum height=7.5mm, mycomponent]
 at ($(objc.east) + (0.4, 0)$) {\footnotesize Type:\texttt{str}\\[-0.4em]\footnotesize Addr:\texttt{0xddd0}\\[-0.4em]\footnotesize Val:\texttt{'d'}};
    \node(objls) [draw=black, fill = white, anchor=south, align=left, minimum width = 12mm, minimum height=7.5mm, mycomponent]
 at ($(objb.north) + (0, 0.3)$) {\footnotesize Type:\texttt{Series}\\[-0.4em]\footnotesize Addr:\texttt{0xeee0}\\[-0.4em]\footnotesize Child:};
     \node(objobj) [draw=black, fill = white, anchor=south, align=left, minimum width = 12mm, minimum height=7.5mm, mycomponent]
 at ($(objc.north) + (0, 0.3)$) {\footnotesize Type:\texttt{MyObj}\\[-0.4em]\footnotesize Addr:\texttt{0xfff0}\\[-0.4em]\footnotesize Child:};

 \node(box1) [draw=black, anchor=south east, minimum width = 2mm, minimum height=1.5mm, inner sep = 0mm]
at ($(objls.south east) + (-0.1, 0.08)$) {};
 \node(box2) [draw=black, anchor=east, minimum width = 2mm, minimum height=1.5mm, inner sep = 0mm]
at ($(box1.west)$) {};
 \node(box3) [draw=black, anchor=east, minimum width = 2mm, minimum height=1.5mm, inner sep = 0mm]
at ($(box2.west)$) {};

 \node(box4) [draw=black, anchor=south east, minimum width = 2mm, minimum height=1.5mm, inner sep = 0mm]
at ($(objobj.south east) + (-0.2, 0.08)$) {};
 \node(box5) [draw=black, anchor=east, minimum width = 2mm, minimum height=1.5mm, inner sep = 0mm]
at ($(box4.west)$) {};

 \node(lstxt) [anchor=east, minimum height=3mm, inner sep = 0.2mm]
at ($(objls.west) + (-0.5, 0)$) {\textbf{ser}};
 \node(objtxt) [anchor=west, minimum height=3mm, inner sep = 0.2mm]
at ($(objobj.east) + (0.5, 0)$) {\textbf{obj}};

\draw[->, thick] 
($(box3.center)$) --
($(obja.north)$); 
\draw[->, thick] 
($(box2.center)$) --
($(objb.north)$); 
\draw[->, thick] 
($(box1.center)$) --
($(objc.north)$); 
\draw[->, thick] 
($(box5.center)$) --
($(objb.north)$); 
\draw[->, thick] 
($(box4.center)$) --
($(objd.north)$); 

\draw[->, thick] 
($(objobj.east) + (0.4, 0)$) --
($(objobj.east) + (0.1, 0)$); 
\draw[->, thick] 
($(objls.west) + (-0.4, 0)$) --
($(objls.west) + (-0.1, 0)$); 

\end{tikzpicture}
\end{subfigure}
\caption{\vargraphs of \texttt{ser} and \texttt{obj} intersect from shared reference to \texttt{'b'} (blue), hence \texttt{\{ser,obj\}} is a Co-variable.}
\label{fig:checkpoint_id_graph}
\end{figure}
\paragraph{Detecting Co-variable updates} 
Co-variable updates is determined by comparing \vargraphs before and after cell executions. A graph structure modification and/or a node attribute change (e.g., object memory address or type) indicates an update to the Co-variable.

\paragraph{Accuracy Guarantee}
As \system constructs \vargraphs following object rechability, it detects Co-variable updates with no false negatives (empirically verified in \cref{sec:experiment_fast_capture_accuracy}). 
However, \system's update detection is \textit{conservative}: there may be \textit{false positives} if objects are dynamically generated (e.g., datatype objects) with a different memory address each time during \vargraph construction/object traversal, or cannot be traversed into (i.e., lacking referencing instructions, e.g., generators~\cite{generators}, which \system assumes to be updated on access).



\subsection{Efficient State Delta Detection}
\label{sec:checkpoint_fast}

This section describes how \system speeds up Co-variable update detection. Identifying updates across the entire global namespace via \vargraphs can be expensive (due to object traversals); \system needs to reduce the number of Co-variables (i.e., portion of namespace) it checks after cell executions without reducing detection accuracy.

\paragraph{Identifying Possibly Updated Co-variables}
Cell executions in Jupyter Notebook interact with the global namespace (i.e., \texttt{globals()}). Therefore, if \system can capture variable references in the cell execution, it can reason about which Co-variables were possibly updated (and which ones were definitely not), as follows:

\begin{definition}
    A Co-variable $\mathcal{X} = \{x_1,...,x_i\}$ is \textbf{accessed} by a cell execution if any variable $x_1,...,x_i$ is accessed (via getting, setting, or deletion) during the cell execution.
\end{definition}

Co-variable accesses indicates \textit{possible} updates (e.g., via a \textit{member function call} \texttt{ser.replace}). \system patches the accessor, setter, and deletion methods of the global namespace (\cref{fig:checkpoint_detect_fast}) to capture variable (hence Co-variable) accesses, which helps identify the \textit{possibly updated Co-variables}: if the members of a Co-variable $\mathcal{X}$ overlaps with the cell execution's accessed variables, then it may have been updated (e.g., \texttt{\{ser,obj\}} in \cref{fig:checkpoint_detect_fast}): \system will verify the update by (1) re-generating \vargraphs for its member variables, (2) comparing the \vargraphs with those before the cell execution to identify modifications, and (3) intersecting the \vargraphs amongst variables of accessed Co-variables to identify merges and splits. Otherwise, the Co-variable surely wasn't updated and \system skips its check for this cell execution (e.g., \texttt{\{df\}} in \cref{fig:checkpoint_detect_fast}, greyed out). 
\begin{figure}[t]

\begin{subfigure}[b]{\linewidth}
\centering
\begin{tikzpicture}[>={LaTeX[width=1mm,length=1mm]},->]

\node(cell) [draw=black, anchor=north, minimum width = 42mm, minimum height=4mm]
at (0,0) {};
\node(codetxt) [anchor=west, minimum height=3mm, inner sep = 0mm,align=left]
at ($(cell.west) + (0.05, 0)$) {\footnotesize\texttt{ser.replace(\textcolor{purple}{'c'},\textcolor{purple}{'e'},inplace=\textcolor{GreenColor}{True})}};

\node(ns) [draw=black, anchor=west, minimum width = 30mm, minimum height=4mm]
at ($(cell.east) + (1.0, 0)$) {};
\node(nstext) [anchor=south west, minimum height=3mm, inner sep = 0mm]
at ($(ns.north west) + (0, 0)$) {\small\textbf{Patched Namespace}};
\node(nsdetecttxt) [anchor=west, minimum height=3mm, inner sep = 0mm,align=left]
at ($(ns.west) + (0.05, 0)$) {\footnotesize\texttt{\_\_getitem\_\_: ls}};

  \node(obja) [draw=purple, thick, anchor=west, minimum width = 4mm, minimum height=4mm, circle, inner sep = 0mm]
 at ($(ns.south) + (-4.2, -0.95)$) {\footnotesize\textcolor{purple}{\texttt{'e'}}};
   \node(objb) [draw=black, thick, anchor=west, minimum width = 4mm, minimum height=4mm, circle, inner sep = 0mm]
 at ($(obja.east) + (0.4, 0)$) {\footnotesize\texttt{'b'}};
   \node(objc) [draw=black, thick, anchor=west, minimum width = 4mm, minimum height=4mm, circle, inner sep = 0mm]
 at ($(objb.east) + (0.4, 0)$) {\footnotesize\texttt{'c'}};
   \node(objd) [draw=black, thick, anchor=west, minimum width = 4mm, minimum height=4mm, circle, inner sep = 0mm]
 at ($(objc.east) + (0.4, 0)$) {\footnotesize\texttt{'d'}};
    \node(objls) [draw=black, thick, anchor=south, minimum width = 4mm, minimum height=4mm, circle, inner sep = 0mm]
 at ($(objb.north) + (0, 0.2)$) {\scriptsize\texttt{ser}};
     \node(objobj) [draw=black, thick, anchor=south, minimum width = 4mm, minimum height=4mm, circle, inner sep = 0mm]
 at ($(objc.north) + (0, 0.2)$) {\scriptsize\texttt{obj}};

      \node(objarr) [draw=black, opacity = 0.3, thick, anchor=east, minimum width = 4mm, minimum height=4mm, circle, inner sep = 0mm]
 at ($(obja.west) + (-0.4, 0)$) {\scriptsize\texttt{arr}};
 \node(objdf) [draw=black, opacity = 0.3, thick, anchor=south, minimum width = 4mm, minimum height=4mm, circle, inner sep = 0mm]
 at ($(objarr.north) + (-0.425, 0.2)$) {\footnotesize\texttt{df}};
 \node(objidx) [draw=black, opacity = 0.3, thick, anchor=east, minimum width = 4mm, minimum height=4mm, circle, inner sep = 0mm]
 at ($(objarr.west) + (-0.4, 0)$) {\scriptsize\texttt{idx}};

      \node(objfit) [draw=black, opacity = 0.3, thick, anchor=west, minimum width = 4mm, minimum height=4mm, circle, inner sep = 0mm]
 at ($(objd.east) + (0.4, 0.35)$) {\scriptsize\texttt{fit}};
       \node(objax) [draw=black, opacity = 0.3, thick, anchor=west, minimum width = 4mm, minimum height=4mm, circle, inner sep = 0mm]
 at ($(objfit.east) + (0.4, -0.35)$) {\tiny\texttt{ax}};
        \node(objfig) [draw=black, opacity = 0.3, thick, anchor=south, minimum width = 4mm, minimum height=4mm, circle, inner sep = 0mm]
 at ($(objax.north) + (0, 0.2)$) {\tiny\texttt{fig}};

\draw[->, thick] 
($(cell.east) + (0.2, 0)$) --
($(ns.west) + (-0.2, 0)$);

\draw[->, thick] 
($(objls.south)$) --
($(obja.north)$); 
\draw[->, thick] 
($(objls.south)$) --
($(objb.north)$); 
\draw[->, thick] 
($(objls.south)$) --
($(objc.north)$); 
\draw[->, thick] 
($(objobj.south)$) --
($(objb.north)$); 
\draw[->, thick] 
($(objobj.south)$) --
($(objd.north)$); 

\draw[->, opacity = 0.3, thick] 
($(objdf.south)$) --
($(objidx.north)$); 
\draw[->, opacity = 0.3, thick] 
($(objdf.south)$) --
($(objarr.north)$); 
\draw[->, opacity = 0.3, thick] 
($(objfig.south)$) --
($(objax.north)$); 

\node[opacity=0.3] (nsdots) at ($(objidx.west) + (-0.4, 0.4)$) {\huge{\bf ...}};
\node[opacity=0.3] (nsdots2) at ($(nsdots.east) + (7.0, 0)$) {\huge{\bf ...}};

\end{tikzpicture}
\end{subfigure}
\caption{\systemnosf efficiently captures state delta via Patched Namespace: it only needs to check Co-variable \texttt{\{ser,obj\}} for updates, as other Co-variables surely weren't updated.}
\label{fig:checkpoint_detect_fast}
\end{figure}
\begin{lemma}
A Co-variable $\mathcal{X} = \{x_1,...,x_i\}$ can be updated by a cell execution only if at least one of $x_1,...,x_i$ was accessed in the code.
\end{lemma}
\begin{proof}
Suppose not, i.e., Co-variable $\mathcal{X}$ does not intersect the accessed variables and was updated. Then, $\mathcal{X}$ must have been updated through via variable $y$ that was not part of $\mathcal{X}$ before the start of the cell execution. Due to Co-variables' self-containment (\cref{sec:checkpoint_blob}), objects reachable from $\mathcal{X}$ cannot possibly be accesed via $y$ during the cell execution without creating a reference by using one of $x_1,...,x_i$ first (e.g., $y.foo$ = $x\_i$), but doing so violates our assumption.
\end{proof}

As only a small portion of variables are accessed per cell in a typical data science notebook, \system significantly reduces delta detection overhead with this approach (empirically verified in \cref{sec:experiment_fast_capture}).

\paragraph{Remark}
As \system patches the notebook session's global namespace, it is impossible for users to use variables from within the notebook (e.g., to modify objects) undetected. Hence, \system will not misidentify Co-variables possibly updated via references.
Users may still use non-referencing methods to update data such as C-pointer-based modifications, but these cases are rare in notebooks (found in 0/60 surveyed notebooks~\cite{lielasticnotebook})
Some libraries do perform memory-based updates (e.g. NumPy's slicing~\cite{numpy}). However, the objects are supported by \system as these updates are still invoked via referencing (e.g., \texttt{arr[0,1] += 1}, empirically verified in \cref{sec:experiment_fast_capture_accuracy}).

\section{Inc. Checkpoint \& Checkout}
\label{sec:restore}
This section describes \system's efficient time-traveling with the Co-variable granularity state deltas. We describe \system' incremental checkpointing in \cref{sec:checkpoint_manage}, \system's incremental checkout in \Cref{sec:restore_fast},
and how \system time-travels to and from notebook states with problematic (e.g., unserializable) data in a fault-tolerant manner in \Cref{sec:restore_robust}.

\subsection{Incremental Checkpointing}
\label{sec:checkpoint_manage}
\begin{figure}[t]

\tikzset{
mylabel/.style={
    font=\footnotesize\sffamily\bfseries,
    align=center,
},
mylabel2/.style={
    font=\footnotesize\sffamily,
    align=center,
},
mycomponent/.style={
    thick, rounded corners=0.5mm,
}
}

\centering
\begin{tikzpicture}[>={LaTeX[width=1mm,length=1mm]},->]




\node(cell2) [draw=black, fill = white, anchor=south, minimum width = 33mm, minimum height=14.3mm, mycomponent]
at (0, 0)  {};
\node(celltext2) [anchor=north west, minimum height=3mm, inner sep = 0.2mm]
at ($(cell2.north west) + (0.1, -0.07)$) {\small CE $t_1$};

\node(cellborder2) [draw=black, anchor=south, minimum width = 31mm, minimum height=6mm]
at ($(cell2.south) + (0, 0.1)$) {};
\node(codetxt2) [anchor=west, minimum height=3mm, inner sep = 0.6mm, align=left]
at ($(cellborder2.west) + (0.05, 0)$) {\footnotesize \texttt{df=load\_csv('...')}\\[-0.35em]\footnotesize \texttt{gmm.init(df)}};

  \node(celldelta2) [draw=black, rotate=90, thick, anchor=west, minimum width = 2.82mm, minimum height=3.5mm, isosceles triangle , inner sep = 0.4mm]
 at ($(cell2.west) + (0.3, 0.005)$) {};
  \node(cellcolons2) [thick, anchor=west, minimum width = 2.82mm, minimum height=4mm, inner sep = 0.4mm]
 at ($(cell2.west) + (0.5, 0.17)$) {\huge{:}};
\node(celldata2) [draw=black, fill = Lightgrey, anchor=west, minimum height=3mm, inner sep = 0.6mm]
at ($(cell2.west) + (0.8, 0.19)$) {\footnotesize (\texttt{\{df\}}, $t_1$)};
\node(celldata22) [draw=black, fill = Lightgrey, anchor=west, minimum height=3mm, inner sep = 0.6mm]
at ($(celldata2.east) + (0.1, 0)$) {\footnotesize (\texttt{\{gmm\}}, $t_1$)};

\node(cell3) [draw=black, fill = white, anchor=south, minimum width = 25mm, minimum height=11.8mm, mycomponent]
at ($(cell2.north) + (-1.4, 0.25)$)  {};
\node(celltext3) [anchor=north west, minimum height=3mm, inner sep = 0.2mm]
at ($(cell3.north west) + (0.1, -0.07)$) {\small CE $t_2$};

\node(cellborder3) [draw=black, anchor=south, minimum width = 23mm, minimum height=3.5mm]
at ($(cell3.south) + (0, 0.1)$) {};
\node(codetxt3) [anchor=west, minimum height=3mm, inner sep = 0.6mm]
at ($(cellborder3.west) + (0.05, 0)$) {\footnotesize \texttt{gmm.fit(k=3)}};

  \node(celldelta3) [draw=black, rotate=90, thick, anchor=west, minimum width = 2.82mm, minimum height=3.5mm, isosceles triangle , inner sep = 0.4mm]
 at ($(cell3.west) + (0.3, -0.115)$) {};
  \node(cellcolons3) [thick, anchor=west, minimum width = 2.82mm, minimum height=4mm, inner sep = 0.4mm]
 at ($(cell3.west) + (0.5, 0.05)$) {\huge{:}};
\node(celldata3) [draw=black, fill = Lightgrey, anchor=west, minimum height=3mm, inner sep = 0.6mm]
at ($(cell3.west) + (0.8, 0.07)$) {\footnotesize (\texttt{\{gmm\}}, $t_{2}$)};

\node(cell4) [draw=BlueColor, fill = white, anchor=south, minimum width = 25mm, minimum height=11.8mm, mycomponent]
at ($(cell3.north) + (0, 0.25)$)  {};
\node(celltext4) [anchor=north west, minimum height=3mm, inner sep = 0.2mm]
at ($(cell4.north west) + (0.1, -0.07)$) {\small CE $t_3$};

\node(cellborder4) [draw=black, anchor=south, minimum width = 23mm, minimum height=3.5mm]
at ($(cell4.south) + (0, 0.1)$) {};
\node(codetxt4) [anchor=west, minimum height=3mm, inner sep = 0.6mm]
at ($(cellborder4.west) + (0.05, 0)$) {\footnotesize \texttt{plot=gmm.result()}};

  \node(celldelta4) [draw=black, rotate=90, thick, anchor=west, minimum width = 2.82mm, minimum height=3.5mm, isosceles triangle , inner sep = 0.4mm]
 at ($(cell4.west) + (0.3, -0.115)$) {};
  \node(cellcolons4) [thick, anchor=west, minimum width = 2.82mm, minimum height=4mm, inner sep = 0.4mm]
 at ($(cell4.west) + (0.5, 0.05)$) {\huge{:}};
\node(celldata4) [draw=black, fill = Lightgrey, anchor=west, minimum height=3mm, inner sep = 0.6mm]
at ($(cell4.west) + (0.8, 0.07)$) {\footnotesize (\texttt{\{plot\}}, $t_{3}$)};

\node(cell5) [draw=black, fill = white, anchor=south, minimum width = 25mm, minimum height=11.8mm, mycomponent]
at ($(cell2.north) + (1.4, 0.25)$)  {};
\node(celltext5) [anchor=north west, minimum height=3mm, inner sep = 0.2mm]
at ($(cell5.north west) + (0.1, -0.07)$) {\small CE $t_4$};

\node(cellborder5) [draw=black, anchor=south, minimum width = 23mm, minimum height=3.5mm]
at ($(cell5.south) + (0, 0.1)$) {};
\node(codetxt5) [anchor=west, minimum height=3mm, inner sep = 0.6mm]
at ($(cellborder5.west) + (0.05, 0)$) {\footnotesize \texttt{gmm.fit(k=10)}};

  \node(celldelta5) [draw=black, rotate=90, thick, anchor=west, minimum width = 2.82mm, minimum height=3.5mm, isosceles triangle , inner sep = 0.4mm]
 at ($(cell5.west) + (0.3, -0.115)$) {};
  \node(cellcolons5) [thick, anchor=west, minimum width = 2.82mm, minimum height=4mm, inner sep = 0.4mm]
 at ($(cell5.west) + (0.5, 0.05)$) {\huge{:}};
\node(celldata5) [draw=black, fill = Lightgrey, anchor=west, minimum height=3mm, inner sep = 0.6mm]
at ($(cell5.west) + (0.8, 0.07)$) {\footnotesize (\texttt{\{gmm\}}, $t_4$)};

\node(cell6) [draw=black, fill = white, anchor=south, minimum width = 25mm, minimum height=11.8mm, mycomponent]
at ($(cell5.north) + (0, 0.25)$)  {};
\node(celltext6) [anchor=north west, minimum height=3mm, inner sep = 0.2mm]
at ($(cell6.north west) + (0.1, -0.07)$) {\small CE $t_5$ (\textbf{\texttt{HEAD}})};

\node(cellborder6) [draw=black, anchor=south, minimum width = 23mm, minimum height=3.5mm]
at ($(cell6.south) + (0, 0.1)$) {};
\node(codetxt4) [anchor=west, minimum height=3mm, inner sep = 0.6mm]
at ($(cellborder6.west) + (0.05, 0)$) {\footnotesize \texttt{plot=gmm.result()}};

  \node(celldelta6) [draw=black, rotate=90, thick, anchor=west, minimum width = 2.82mm, minimum height=3.5mm, isosceles triangle , inner sep = 0.4mm]
 at ($(cell6.west) + (0.3, -0.115)$) {};
  \node(cellcolons6) [thick, anchor=west, minimum width = 2.82mm, minimum height=4mm, inner sep = 0.4mm]
 at ($(cell6.west) + (0.5, 0.05)$) {\huge{:}};
\node(celldata6) [draw=black, fill = Lightgrey, anchor=west, minimum height=3mm, inner sep = 0.6mm]
at ($(cell6.west) + (0.8, 0.07)$) {\footnotesize (\texttt{\{plot\}}, $t_5$)};

\draw[->, ultra thick] 
($(cell2.north)$) --
($(cell3.south)$); 
\draw[->, ultra thick] 
($(cell2.north)$) --
($(cell5.south)$); 
\draw[->, ultra thick] 
($(cell5.north)$) --
($(cell6.south)$); 
\draw[->, ultra thick] 
($(cell3.north)$) --
($(cell4.south)$); 

\draw[->, ultra thick, densely dashed, opacity=0.6]
($(celldata3.north)$) --
($(cellborder4.south) + (-0.2, 0)$); 

\node(step1) [anchor=center, minimum height=6mm, inner sep = 0.2mm,align=center]
at ($(cell4.west) + (-1.4, 0)$) {\footnotesize \textbf{(1)}: State delta of\\[-0.35em]\footnotesize updated Co-variables};

\node(step2) [anchor=center, minimum height=6mm, inner sep = 0.2mm,align=center]
at ($(cell4.west) + (-1.4, -1)$) {\footnotesize \textbf{(2)}: Cell code of\\[-0.35em]\footnotesize execution};

\node(step2) [anchor=center, minimum height=6mm, inner sep = 0.2mm,align=center]
at ($(cell4.west) + (-1.4, -2)$) {\footnotesize \textbf{(3)}: dependencies of\\[-0.35em]\footnotesize execution};

\begin{scope}[->,
              >=stealth',
              every node/.style={fill=none,circle, inner sep=0mm},
              every edge/.style={draw=black}]
    \node[] (y) at ($(celldata4.north west) + (-0.6, 0)$) {};
    \node[] (z) at ($(celldata4.north west) + (-1.3, 0)$) {};
    \node[] (y1) at ($(cellborder4.west) + (0, 0)$) {};
    \node[] (z1) at ($(cellborder4.west) + (-0.7, -0.3)$) {};
    \node[] (y2) at ($(cell4.south) + (-0.2, -0.2)$) {};
    \node[] (z2) at ($(cell4.south) + (-1.6, -1.1)$) {};
    \draw[->, black, thick, densely dotted] (y)  to [bend right = 15] (z);
    \draw[->, black, thick, densely dotted] (y1)  to [bend right = 15] (z1);
    \draw[->, black, thick, densely dotted] (y2)  to [bend right = 25] (z2);
\end{scope}

\end{tikzpicture}
\caption{A Checkpoint Graph with 2 branches ($t_1\rightarrow t_2 \rightarrow t_3$ and $t_1\rightarrow t_4 \rightarrow t_5$). \system manages state deltas in the Checkpoint Graph; Co-variables are versioned by update time.}
\label{fig:checkpoint_graph}
\end{figure}
This section describes how \system performs incremental checkpointing by writing and managing per-cell-execution checkpoints containing the updated Co-variables with the \textit{Checkpoint Graph}.

\paragraph{Checkpoint Graph}

The Checkpoint Graph is a directed tree of (incremental) checkpoints representing the branch-based state evolution.
Nodes are added for each of \system's checkpoints, and are timestamped with the completion time $t$ of the corresponding cell execution (we refer to the timestamped node and cell execution as \textit{node $t$} and \textit{CE $t$}, respectively).
The Checkpoint Graph maintains a \textit{head node} tracking the current state. Each node $t$ contains the state delta consisting of Co-variables updated by CE $t$.
Co-variables stored in each node $t$ are versioned accordingly:

\begin{definition}
    A \textbf{Versioned Co-variable} is a Co-variable-timestamp pair $(\mathcal{X}, t)$ representing the Co-variable $\mathcal{X}$ updated by CE $t$.
\end{definition}

Versioned Co-variables are analogous to versioned datasets: the same Co-variable (w.r.t. variable membership) can take on multiple values during a session being updated by different cell executions. \cref{fig:checkpoint_graph} show an example: CE $t_3$ creates the Co-variable \texttt{\{plot\}}, which is stored in node $t_3$ (red) as the \textit{Versioned Co-variable} $((plot), t_3)$.

\paragraph{Writing into the Checkpoint Graph}
After each CE $t$, \system writes a new node $t$ in the Checkpoint Graph with (1) the versioned Co-variables in CE $t$'s state delta, (2) CE $t$'s code, and (3) CE $t$'s accessed versioned Co-variables stored in previous checkpoints (\cref{sec:checkpoint_fast}).
For example, node $t_3$ in \cref{fig:checkpoint_graph} (blue) contains the code (``\texttt{plot=gmm.result()}``) and its dependency on $((gmm), t_2)$ from node $t_2$  (dashed line). Notably, the state delta, code, and variable accesses are respectively analogous to the \textit{update}, \textit{operation}, and \textit{dependencies} in database logging and versioning.
The new node $t$ is written under the head node $s$, and and a parent-child relation is added from $s$ to $t$ (now the new head node).

\paragraph{Handling Unserializable Data}
If \system cannot write an updated Co-variable into the Checkpoint Graph (e.g., it contains an unserializable object such as a generator~\cite{generators} or hash~\cite{hashlib}), \system simply skips its storage.
Instead, upon checkout, the missing Co-variable will be restored through fallback recomputation enabled by the cell code and dependencies stored in the Checkpoint Graph node (\cref{sec:restore_robust}).

\subsection{Efficient State Restoration}
\label{sec:restore_fast}
\subfile{plots/fig_restore_diff}
\system's goal for incremental checkout is to accurately and efficiently restore the current state to the target state.
To do so, it must identify the contents of the target state via its timestamp, analogous MVCC's timestamped snapshots~\cite{bernstein1983multiversion}; instead of versioned tables, we identify \textit{Versioned Co-variables} in the (timestamped) target state:

\begin{definition}
    The \textbf{Session State} at timestamp $t$  is a set of $n$ Versioned Co-variables $\{(\mathcal{X}_i, t_i)|1 \leq i \leq n\}$ such that for each $(\mathcal{X}_i, t_i)$:\begin{enumerate}
        \item $t_i$ is an ancestor of $t$ on the Checkpoint Graph.
        \item There must not exist another versioned Co-variable $(\mathcal{Y}_j, t_j)$ such that $\mathcal{X}_i \cup \mathcal{Y}_j \neq \emptyset$ and $t_j$ is a child of $t_i$ and ancestor of $t$.
    \end{enumerate}
\end{definition}

The session state at timestamp $t$\footnote{\system stores snapshots of Session State metadata (i.e., references to contained Co-variables) in Checkpoint Graph nodes.} (\textit{state $t$} for brevity) is the set of all Versioned Co-variables that are in the namespace after CE $t$, i.e., not overwritten by a newer Versioned Co-variable prior to CE $t$. For example, in \cref{fig:restore_diff}, state $t_{3}$ (top-left) consists of $(\{plot\}, t_3)$, $(\{gmm\}, t_2)$, and $(\{df\}, t_1)$. It does not contain $(\{gmm\}, t_1)$ as it was overwritten by CE $t_2$ (\texttt{gmm.fit(k=3)}) which writes $(\{gmm\}, t_2)$. 
Each state $t$ dictates which Versioned Co-variables should be loaded from various Checkpoint Graph nodes for checkouts; for efficient incremental checkout, \system identifies the current and target states' \textit{difference} w.r.t. the (versioned) Co-variables that need updating: some Co-variables do not need updating when converting the current state to the target state, identifiable via the Checkpoint Graph:
\begin{definition}
    A Co-variable $\mathcal{X}$ is \textbf{identical} between the current state $t_a$ and target state $t_b$ if a Versioned Co-variable $(\mathcal{X}, t_c)$ exists in the session states of $t_a$, $t_b$, and $t_c$, where $t_c$ is the \textbf{lowest common ancestor} of node $t_a$ and node $t_b$.
    Otherwise, if no such $(\mathcal{X}, t_c)$ exists, then the Co-variable $\mathcal{X}$ has \textbf{diverged} between $t_a$ and $t_b$.
\end{definition}

A Co-variable $\mathcal{X}$ is identical between states $t_a$ and $t_b$ if its versioned counterpart is consistent across $t_a$, $t_b$, and $t_c$, i.e., no CE between (1) nodes $t_a$ and $t_c$ and (2) nodes $t_b$ and $t_c$ updated $\mathcal{X}$, hence does not need updating when checking out from $t_a$ to $t_b$.
For example, in \cref{fig:restore_diff}, if checking out from $t_5$ to $t_3$, the Co-variable \texttt{\{df\}} (blue) is identical between the states as no CE between (1) nodes $t_1$ and $t_3$ (2) nodes $t_1$ and $t_5$ updated it. Otherwise, if the Co-variable $\mathcal{X}$ has diverged between the current and target states, it will need updating (by either loading an appropriate Versioned Co-variable or deleting it) to checkout to the target state.
For example, the Co-variable \texttt{\{gmm\}} (red) has diverged between nodes $t_5$ and $t_3$ as their parents ($t_4$ and $t_2$) both updated \texttt{gmm} with their CE (fitting with \texttt{k=3} and \texttt{k=10}), hence, \texttt{gmm} (and \texttt{plot}) needs updating via loading \texttt{(\{gmm\},$t_2$)} if checking out from state $t_5$ to state $t_3$.
\subfile{plots/fig_restore_fallback}
\paragraph{Performing State Checkout}
When checking out to the state at node $t$, The State Restorer (\cref{sec:system_components}) performs the following steps:\begin{enumerate}
    \item Load the appropriate Versioned Co-variables from nodes (i.e., node $t$ and ancestors of node $t$) to update diverged Co-variables between the state of the current head node $s$ and node $t$.
    \item Update/re-generate \vargraphs (\cref{sec:checkpoint_detect}) for updated Co-variables.
    \item Move the head from node $s$ to the checked out node $t$.
\end{enumerate}
Notably, the next cell execution will create a node in a new branch rooted at $t$ in the Checkpoint Graph, e.g., the graph in \cref{fig:restore_diff} is generated through the sequence $t_1\rightarrow t_2\rightarrow t_3\rightarrow(\text{checkout to }\,  t_1)\rightarrow t_4 \rightarrow t_5$.
If during checkout, a required Versioned Co-variable is missing (i.e., due to serialization failure, \cref{sec:checkpoint_manage}) or fails to load (i.e., deserialization failure), \system restores it via fallback recomputation.

\subsection{Robust Restoration}
\label{sec:restore_robust}

In this section, we describe how \system restores problematic data to achieve generalizable and fault-tolerant incremental checkout.

\paragraph{Fallback Recomputation}
As each Checkpoint Graph node $t$ contains the code of CE $t$ and (2) which previous Versioned Co-variables $(\mathcal{X}_j, t_j)$ CE $t$ accessed (\cref{sec:checkpoint_manage}), any Versioned Co-variable in node $t$'s state delta can be recomputed by (1) loading accessed Versioned Co-variables from previous checkpoints, then re-running CE $t$.
For example, in \cref{fig:restore_fallback}, suppose (\texttt{\{plot\}}, $t_3$) (green) fails to load when checking out to state $t_3$. (\texttt{(\{gmm\}}, $t_2$) is required to rerun CE $t_3$ (red); therefore, (\texttt{\{gmm\}}, $t_2$) is loaded from the parent node $t_2$, and after rerunning CE $t_3$ on the input (\texttt{\{gmm\}}, $t_2$), (\texttt{\{plot\}}, $t_3$) is restored.

\paragraph{Dynamic and Recursive Fallbacks}
\system's fallback recomputation is dynamic and recursive---if another Co-variable is missing or fails to load when retrieving recomputation inputs, fallback recomputation can be recursively performed for that Co-variable.
For example, if (\texttt{\{gmm\}}, $t_2$) from node $t_2$ fails to load (as fallback recomputation for (\texttt{\{plot\}}, $t_3$) from node $t_3$), it itself can be recomputed by loading (\texttt{\{gmm\}}, $t_1$) from node $t_1$ and rerunning CE $t_2$ (blue).

\paragraph{Remark} 
\system guarantees \textit{exact} restoration for all serializable Co-variables w.r.t. \textit{bytestring representation} if there are no hidden serialization errors (\cref{sec:implementation_consideration}).
While \system's fallback recomputation restores some problematic Co-variables, exactly restoring Co-variables that both (1) fail to store/load and (2) are created non-deterministically (e.g., random generators) is currently unsupported.
This limitation is similarly seen in Spark~\cite{zaharia2010spark} and Ray's~\cite{moritz2018ray} lineage-based fault tolerance; however, in our case, such objects are rare in data science libraries (\cref{sec:experiments_robust}), hence we consider this limitation to be acceptable.

\section{Implementation and Discussion}
\label{sec:implementation}

This section describes \system's implementation details (\Cref{sec:implementation_details}) and design considerations (\Cref{sec:implementation_consideration}).

\subsection{Implementation}
\label{sec:implementation_details}

\paragraph{Integrating with Jupyter} \system is implemented as a separate application from the notebook process, usable without altering the base Jupyter application. Upon session initialization, \system places hooks into the kernel (\texttt{pre\_run\_cell} and \texttt{post\_run\_cell}~\cite{ipythonhook}) and patches the namespace (\texttt{user\_ns}~\cite{ipythonns}) (\cref{sec:system_components}), allowing the standalone \system process to detect state deltas, write data to storage, and overwrite data in the namespace upon checkout transparently.

\paragraph{Serialization Protocol}
The Pickle protocol (i.e., \texttt{\_\_reduce\_\_}~\cite{pickle}) is 
used
for (1) object serialization and (2) constructing \vargraphs for identifying Co-variables, i.e., object \texttt{y} is reachable from another object \texttt{x} if \texttt{pickle(x)} includes \texttt{y}.
As Pickle is the de-facto standard (in Python) observed by almost all libraries (e.g., NumPy, PyTorch~\cite{pytorchcheckpoint}), \system can be used in almost all cases. \system's per-Co-variable storage also enables \textit{mixing and matching} serialization libraries for coverage:
Currently, \system will try CloudPickle~\cite{cloudpickle} first, then use Dill~\cite{dill} as a fallback for Co-variables that CloudPickle fails on.
\paragraph{Storing Checkpoints}
\system uses SQLite~\cite{sqlite} to store Versioned Co-variables in the Checkpoint Graph. However, any storage mechanism~\cite{chockchowwat2022airphant, chockchowwat2022automatically, chockchowwat2023airindex} can be used in its place---even in-memory ones if the user wants to maximize checkpointing/checkout efficiency.

\subsection{Design Considerations}
\label{sec:implementation_consideration}




\paragraph{Silent Serialization Errors}
Certain object classes may contain incorrect serialization instructions, which, despite being able to be stored/loaded to/from storage, result in silent errors.
\system currently assumes that instructions are correctly implemented for all objects w.r.t. equality before and after pickling, and does not prevent these silent errors.
However, these cases are rare (\cref{sec:experiment_fast_capture_accuracy}), and \system provides a blocklist file for users to force fallback recomputation for Co-variables containing objects belonging to these classes.
\begin{table}[t]

\caption{Summary of Notebooks for Evaluation.}
\footnotesize
\addtolength{\tabcolsep}{-3.5pt} 
\midsepremove
\begin{tabular}{lllrrrr}
\toprule
\textbf{Notebook} & \textbf{Topic} & \textbf{Library} & \textbf{Cells} & \textbf{Time(s)} & \textbf{Data(MB)} & \textbf{Final} \\
 \midrule
Cluster\cite{bruteforce}&  Cluster analysis & seaborn\cite{seaborn} & 24 & 1703 & 43 & Yes   \\
TPS\cite{tpsmar} & Random forest & intelex\cite{scikitlearnintelex} & 49 & 154 & 31 & Yes \\
Sklearn\cite{sklearntweet}& Text mining & sklearn\cite{scikitlearn} & 44 & 512 & 185 & No  \\
HW-LM\cite{hwlm}& Linear regression & NumPy\cite{numpy} & 81 & 13 & 1 & Yes\\
StoreSales\cite{storesales}& TS analysis & SM\cite{statsmodels} & 41 & 665 & 122 & Yes\\
Qiskit\cite{qiskit} & Quant. Computing & Qiskit\cite{qiskitrepo} & 85 & 46 & 1 & No \\
TorchGPU\cite{pytorchresnet} & Image classification & PyTorch\cite{pytorch} & 27 & 716 & 1090 & Yes\\ 
Ray\cite{raynb} & Distrib. Computing & Ray\cite{rayrepo} & 20 & 2361 & 92 & No\\
\bottomrule
\end{tabular}
\midsepdefault
\addtolength{\tabcolsep}{3.5pt}
\label{tbl:workload}
\end{table}

\paragraph{Alternative Delta Detection Methods}
While \system's \vargraphs can generalizably detect Co-Variable updates, there are specific cases that allow for more efficient detection methods such as (1) hashing (e.g., XXH64~\cite{xxhash}) for array-likes~\cite{numpy}) and (2) rule-based static cell (e.g., \texttt{df.head}) identification for skipping update detection (\cref{sec:motivation_workload}). \system currently uses hashing for common array-likes (e.g., tensors~\cite{tensorflow}), but can be extended to incorporate (1) other detection methods for specific classes and (2) rule-based cell analyses.

\section{Experimental Evaluation}
\label{sec:experiments}
In this section, we empirically study the effectiveness of \system's time-traveling.
We make the following claims: 
\begin{enumerate}
\item \textbf{Generalizable and Robust Mechanism:} \system can identify modifications to, and correctly restore session states containing 146 object classes from common Data Science libraries. (\Cref{sec:experiments_robust})
\item \textbf{Low Checkpoint Storage Cost:} \system's optimizations result in its per-cell-execution checkpoints being up to 4.55$\times$ smaller compared to those from the next best mechanism. (\Cref{sec:experiment_cheap_checkpoint})
\item \textbf{Low Checkpoint Times:} \system's checkpoints are created up to 5.12$\times$ faster compared to the next best mechanism. (\Cref{sec:experiment_fast_checkpoint})
\item \textbf{Fast Incremental Checkout:} \system's novel incremental restoration is crucial to its \textit{sub-second} checkout times --- up to 8.18$\times$ and 4.18$\times$ faster than the next best mechanism for undoing cell executions and switching branches, respectively. (\Cref{sec:experiment_checkout})
\item \textbf{Low Overhead Delta Detection:} \system incurs negligible runtime overheads on data science notebooks for capturing the state delta --- less than 3.0\% of the notebook session runtime and up to 4.08$\times$ less than alternative tracking approaches. 
(\Cref{sec:experiment_fast_capture})
\end{enumerate}
\begin{table}[t]

\caption{Categorization of 146 Object Classes for Evaluation. Referred to in \cref{fig:experiment_robust}, \cref{tbl:experiment_robust_example_table}, and \cref{tbl:experiment_fast_capture_accuracy}.}
\footnotesize
\addtolength{\tabcolsep}{-2pt} 
\midsepremove
\begin{tabular}{lll}
\toprule
\textbf{Category} & \textbf{Example Libraries} & \textbf{Example Class} \\
 \midrule
Data Analysis & pandas\cite{pandas}, polars\cite{polars}, pyarrow\cite{pyarrow} & pd.DataFrame\cite{pddataframe} \\
Data Visualization & matplotlib\cite{matplotlib}, plotly\cite{plotly}, seaborn\cite{seaborn} & plt.Figure\cite{pltfigure} \\
Machine Learning & sklearn\cite{scikitlearn}, xgboost\cite{xgboost}, scipy\cite{scipy} & GMM\cite{sklearnmixture} \\
Deep Learning & tensorflow\cite{tensorflow}, torch\cite{pytorch}, keras\cite{keras} & torch.Tensor\cite{torchtensor}\\
NLP & nltk\cite{nltk}, textblob\cite{textblob}, wordcloud\cite{wordcloud} & TextBlob\cite{textblob} \\
Computer Vision & photutils\cite{photutils}, torchvision\cite{torchvision} & ImageDepth\cite{imagedepth} \\
Dist. Computing & pyspark\cite{pyspark}, ray\cite{rayrepo}, optuna\cite{optunarepo} & pyspark.sql\cite{pysparksql} \\
Data Pipelining & huggingface\cite{huggingfacerepo}, transformers\cite{transformersrepo} & BertTokenizer\cite{hfberttokenizer} \\
\bottomrule
\end{tabular}
\midsepdefault
\addtolength{\tabcolsep}{2pt}
\label{tbl:library}
\end{table}
\subsection{Experiment Setup}
\label{sec:exp_setup}

\paragraph{Datasets} 
We select 8 data science notebooks from Kaggle Grandmaster-level users or Github-hosted tool tutorials (e.g., Ray) (\cref{tbl:workload}), each featuring a popular data science library, which we categorize as \textit{in-progress} (3/8) or \textit{final} (5/8), with the former containing out-of-order cell executions and the latter lacking them. Notably, the final notebooks are also cleaned and contain memory/runtime optimizations. We empirically verify all notebooks follow traits discussed in \cref{sec:motivation_workload}, and provide more categorization details in the appendix (\cref{sec:appendix}).
%

We also select 146 common data science library classes ( \cref{tbl:library}), on which we evaluate \system's correctness and robustness.

\begin{figure}[t]
\usetikzlibrary{patterns}
\begin{subfigure}[b]{\linewidth}
\centering
\begin{tikzpicture}

\pgfplotstableread[col sep=comma,]{
name
\criu
\dumpsession
\system \textbf{(Ours)}
}\datatable

\begin{axis}[
    ybar,
    clip=false,
    xlabel style={yshift = 4ex},
    width=85mm,
    height=26mm,
    bar width=5mm,
    ymin=0,
    ymax=8,
    ylabel style={yshift = -3ex},
    axis y line*=none,
    axis x line*=none,
    ytick={0, 2, 4, 6, 8},
    yticklabels={0, 2, 4, 6, 8},
    xtick={1, 2, 3},
    xtick style ={draw=none},
    xticklabels={\criu, \dumpsession, \system \textbf{(Ours)}},
    x tick label style={yshift = 1ex},
    xmin=0.5,
    xmax = 3.5,
    ymajorgrids,
    tick label style={font=\footnotesize},
    legend style={
        font=\footnotesize,
        /tikz/every even column/.append style={column sep=0.5cm},
        legend columns = 3,
        at={(-0.15, 1.1)}, anchor=south west
    },
    label style={font=\footnotesize},
    ylabel={Failures},
    area legend
    ]

    \addplot[NoOptColor,fill=NoOptColor]coordinates {(1.235,6)};
      \addplot[LRUColor,fill=LRUColor]coordinates{(2, 7)};
      \addplot[black, dashed, thick, fill=none]coordinates{(2.765,8)};
                 
    


    \node[anchor=south west, GreenColor, rotate=90,
        font=\scriptsize\bfseries] 
        at (axis cs: 3.09, -0.4) 
        {0 failures};


\end{axis}
    
    
\end{tikzpicture}
\end{subfigure}

\caption{Checkpoint/checkout failures. 
\systemnosf successfully checkpoints/checkouts all object classes with no failures.
}
\label{fig:experiment_robust}
\end{figure}



\paragraph{Methods} We evaluate \system against existing tools capable of enabling time-travelling on notebooks to various degrees:
\begin{itemize}
    \item \criu~\cite{criu}: Performs a system-level memory dump of the process hosting the notebook session. 
        The session state is restored by loading the memory dump and reviving the process.
    \item \criuincremental~\cite{criu}: \criu with snapshot deduplication, storing only dirty memory pages in subsequent snapshots.
    \item \dumpsession~\cite{dumpsession}: An application-level checkpointing tool that serializes the entire session state into one single file.
    \item \elasticnotebook~\cite{lielasticnotebook}: An application-level notebook migration tool that balances data serialization and cell recomputation to achieve optimized session replication times.
    \item \detreplay: A checkpoint-optimized version of \system based on operation replay~\cite{mohan1992aries, netezzanondet} that uses manual annotation\footnote{Automatically detecting non-/determinism in executions is out of scope for this work.} to skip incremental checkpointing for deterministic cells. These deterministic cells are replayed as necessary on checkout.
    
\end{itemize}

\paragraph{Ablation Study} We additionally compare the overhead of \system's update detection mechanism with these tracking methods (\Cref{sec:experiment_fast_capture}):\begin{itemize}
    \item \ipyflow~\cite{shankar2022bolt}: A hybrid dynamic-static (i.e., AST analysis with live symbol resolution) for obtaining sub-variable (i.e., symbols, e.g., \texttt{ls[x]}) level granularity to perform reactive cell executions.
    \item \systembad: Always perform update detection for all Co-variables in the session state after each cell execution, regardless of whether they were accessed in the previous cell.
\end{itemize}


\paragraph{Methodology and Measurement} We run notebook cells sequentially from top to bottom and checkpoint after each cell execution. We checkout into the same state for \system and \detreplay and into a fresh kernel/process for other methods. We measure the (1) \emph{checkpoint time} (including both tracking and data writing) after each cell execution, (2) \emph{checkout time} to restore the state from checkpoint files, and (3) \emph{tracking overhead} of \system after each cell execution to track updates. 
We clear the page cache between runs.
\paragraph{Environment and Reproducibility}
Experiments are performed on an Ubuntu server with 2 AMD EPYC 7552 48-Core Processors and 1TB RAM. All checkpoints are written to a mounted NFS, with disk read and write speeds of 519.8 MB/s and 358.9 MB/s, respectively. Our Github repository\footnote{\url{https://github.com/illinoisdata/kishu-vldb}} contains our implementation of \system, experiment notebooks/library classes, and scripts.

\subsection{Generalized and Robust Time Traveling}
\label{sec:experiments_robust}
\begin{table}[t]
\caption{\systemnosf handles these classes existing works fail on.}
\label{tbl:experiment_robust_example_table}
\footnotesize
\addtolength{\tabcolsep}{-3pt} 
\midsepremove
\begin{tabular}{lll}
\toprule
 \textbf{Tool} & \textbf{Description} & \textbf{Failure Classes}\\
\midrule
CRIU & Dist. Computing & pyspark.sql~\cite{pysparksql}, ray.Dataset~\cite{raydataset} \\
 & On-device data & tf.tensor~\cite{tftensor}, torch.tensor~\cite{torchtensor}\\
 & Data Pipelining & Pipeline~\cite{hfpipeline}, BertTokenizer~\cite{hfberttokenizer}\\
 \midrule
\dumpsession & Unserializable Data & pl.LazyFrame~\cite{pllazyframe}, bokeh.figure~\cite{bokehfigure}\\
\bottomrule
\end{tabular}
\midsepdefault
\addtolength{\tabcolsep}{3pt} 

\end{table}

\begin{table}[t]

\caption{Summary of \systemnosf's update detection.}
\footnotesize
\addtolength{\tabcolsep}{-3pt} 
\midsepremove
\begin{tabular}{llr}
\toprule
\textbf{Result} & \textbf{Description} & \textbf{Count}  \\
 \midrule
Success & \system reports an update when object is changed & 120 \\
False Positive & \system reports update on access when object is unchanged& 14 \\
Pickle Error & Object can't be deterministically stored, \system reports update & 12 \\
Fail & Object is changed but \system does not report an update & 0 \\
\bottomrule
\end{tabular}
\midsepdefault
\addtolength{\tabcolsep}{3pt}
\label{tbl:experiment_fast_capture_accuracy}
\end{table}

This section compares the robustness of \system's time-traveling to existing methods. We attempt to checkpoint and checkout session states containing objects from the 146 data science library classes and compare number of classes each method fails to checkout.
\subfile{plots/experiment_cheap_checkpoint}
\subfile{plots/experiment_fast_checkpoint}

We report results in \cref{fig:experiment_robust}. \system completes time-traveling for all 146 libraries, handling 6 classes with multiprocessing and/or off-CPU data and 7 unserializable classes that \criu and \dumpsession fail on, respectively: unlike \criu, \system utilizes reductions (\cref{sec:implementation_details}) to store Co-variables, hence it can store distributed or off-CPU data (e.g., Ray's dataset\cite{raydataset} or on-GPU tensors\cite{tftensor, torchtensor}) and unlike \dumpsession, \system's fallback recomputation allows it to restore Co-variables with (1) unserializable objects (e.g., pl.LazyFrame\cite{pllazyframe}) or (2) serializable objects that can't deserialize (e.g., bokeh.figure\cite{bokehfigure}). \cref{tbl:experiment_robust_example_table} summarizes these noteworthy classes.

\subsubsection{Accurate Delta Detection}
\label{sec:experiment_fast_capture_accuracy}

We verify \system's delta detection accuracy by comparing two \vargraphs generated for each class object before and after (1) updating a class attribute (e.g., \texttt{model.key = 'A'}) or (2) updating nothing. We count the number of \vargraph differences for case (1) as \textit{successes} and case (2) as \textit{false positives}.

We report results in \cref{tbl:experiment_fast_capture_accuracy}. \system's \vargraphs accurately captures object updates in 120 classes. While \system reports false positives in 14 classes, (e.g., due to dynamically generated reachable objects), they only affect \system's efficiency (i.e., during time-traveling); however, \system maintains accuracy by considering these objects to be updated on access. We also find that 12 classes contain \textit{silent pickling errors} (\cref{sec:implementation_consideration}); nevertheless, \system reports these objects to be updated on access similar to false positives, and users may force their (fallback) recomputation if needed (\cref{sec:implementation_consideration}). Notably, \system has no false negatives: \system will always report if an object is changed.

\subsection{Small Incremental Checkpoint Sizes}
\label{sec:experiment_cheap_checkpoint}

This section compares \system's checkpoint sizes with those of existing tools: we checkpoint the session state after each cell execution with each method and measure the total storage size of checkpoints.

We report results in \cref{fig:experiment_cheap_checkpoint}. \system's cumulative checkpoint size is consistently the smallest (expectedly except for \detreplay, explained shortly)
and is up to 4.55$\times$ smaller than the next best alternative (\textit{HW-LM}). 
\elasticnotebook, while the next best method on 6/8 notebooks and also has fault-tolerant mechanisms to checkpoint all 8 notebooks, can fall short in checkpointing time (\cref{sec:checkpoint_fast}).
\criuincremental, while also incrementally checkpointing, is not the next best method on any notebook, losing to \elasticnotebook and \dumpsession on 6 and failing to checkpoint on 2 as it (1) incrementally checkpoints at the coarser memory page level (\cref{sec:motivation_delta}), and (2) does not handle off-CPU data and multiprocessing (\cref{sec:experiments_robust}).
\dumpsession fails on \textit{Qiskit} as it cannot handle unserializable data, and \criu incurs prohibitive storage costs (94GB, \textit{TPS}) as it non-incrementally checkpoints at OS-level.
While \detreplay can save checkpoint storage cost of up to 3.95$\times$ vs. \system (\textit{StoreSales}) by skipping checkpointing after deterministic cells, it (1) needs manual annotation and (2) can result in unacceptable checkout times (\cref{sec:experiment_checkout}).

\subsection{Low Incremental Checkpoint Time}
\label{sec:experiment_fast_checkpoint}

This section compares the checkpoint time of \system with that of existing tools: we measure the total time spent by each method creating checkpoints after each cell execution.

We report results in \cref{fig:experiment_fast_checkpoint}. \system's cumulative checkpointing time is the lowest (except for \detreplay) on 5/8 notebooks, being only up to 15.5\% of notebook runtime (\textit{HW-LM}) and up to 5.12$\times$ faster (\textit{HW-LM}) than the next best alternative on these notebooks. 
While \criuincremental checkpoints faster than \system on 3/8 notebooks owing to memory dumping being faster than serialization for unit data, the difference is negligible (up to 3.03$\times$, \textit{StoreSales}) compared to the reliability issues (\cref{sec:experiments_robust}), space inefficiency (\cref{sec:experiment_cheap_checkpoint}), and slow checkout times (\cref{sec:experiment_checkout}).
Compared to \elasticnotebook, \system's checkpointing is EAFP-based~\cite{goodger2014code}: if it fails to store a Co-variable, it will simply recompute it upon checkout via fallback recomputation. This allows it to skip the profiling (i.e., for data sizes and serializability) required for \elasticnotebook's optimization (for what to store/recompute), which causes checkpoint times slower than \dumpsession on 2/8 notebooks.

\subsection{Fast Incremental Checkout}
\label{sec:experiment_checkout}

This section compares the efficiency of \system's incremental checkout with the (non-incremental) checkout of existing methods. We generate per-cell-execution checkpoints on the notebooks following the methodology in \cref{sec:experiment_cheap_checkpoint} and \cref{sec:experiment_fast_checkpoint}, then measure the time it takes for each method to checkout to a previous state (i.e. undo, \cref{sec:experiment_checkout_undo}) or checkout to a different execution branch (\cref{sec:experiment_checkout_branch}).

\subsubsection{Fast Execution Undo}
\label{sec:experiment_checkout_undo}

For each notebook, we measure the time it takes to undo various dataframe and plot operation cells. 

We report the results in \cref{fig:experiment_checkout_undo}. \system is the only method capable of incrementally checking out: it achieves \textit{sub-second} cell execution rollbacks on all test cases, and is up to 8.18$\times$ faster than the next best alternative (\textit{StoreSales}).
While \criuincremental achieves checkpoint times comparable with \system, it is up to \textbf{36$\times$} slower for checking out (\textit{StoreSales}) and the slowest method for undos on 5/6 notebooks, due to it needing to piece together the memory snapshot of the notebook process to restore from multiple (incremental) checkpoint files.
\criu, \dumpsession, and \elasticnotebook cannot incrementally checkout hence they cannot consistently perform sub-second undos.
For example, the \textit{Sklearn} notebook test case drops a column in an \textit{auxiliary dataframe} that is 1.4MB in size (vs. the 133MB main dataframe).
\system identifies that it only needs to load the auxiliary dataframe from before the cell execution and undoes the operation in 0.4 seconds; however, other non-\detreplay methods all require the entire session state to be overwritten with a complete load of checkpoint data, taking an upwards of 6 seconds to do so (and in \criu and \criuincremental's case, also killing and restarting the current notebook process). 
\begin{figure}[t]
\usetikzlibrary{patterns}
\begin{subfigure}[b]{\linewidth}
\centering
\begin{tikzpicture}

\pgfplotstableread[col sep=comma,]{
name
Cluster (6)
Sklearn (29)
HW-LM (61)
StoreSales (27)
Qiskit (75)
Ray (17)
}\datatable

\begin{axis}[
    ybar,
    clip=false,
    xtick={1, 2, 3, 4, 5, 6},
    xticklabels from table={\datatable}{name},
                 x tick label style={anchor=center, yshift = 0ex, font=\scriptsize},
    xtick style ={draw=none},
    xlabel style={yshift = 2.5ex, font=\footnotesize},
    ylabel style={yshift = -3.5ex, font=\scriptsize},
    width=90mm,
    height=26mm,
    bar width=1.0mm,
    ymin=0.01,
    ymax=100,
    log origin = infty,
    ymode = log,
    axis y line*=none,
    axis x line*=none,
    ytick={0.01, 0.1, 1, 10, 100},
    yticklabels={0.01, 0.1, 1, 10, 100},
    xmin=0.5,
    xmax = 6.5,
    ymajorgrids,
    tick label style={font=\footnotesize},
    legend style={
        font=\footnotesize,
        /tikz/every even column/.append style={column sep=0.2cm},
        legend columns = 3,
        inner ysep=0.5pt,
        at={(-0.06,1.1)},
        anchor=south west,
    },
    ylabel={Checkout Time (s)},
    area legend
    ]

    \addplot [black,fill=NoOptColor,x tick label style={xshift=-0.3cm}, postaction={
        pattern=north west lines
    }] table[x=notebook,y=criu] {sections/data/restore_undo_time_table.txt};
    \addlegendentry[]{\criu};

    \addplot [black,fill=LRUColor,x tick label style={xshift=-0.3cm}, postaction={
        pattern=horizontal lines
    }] table[x=notebook,y=dump] {sections/data/restore_undo_time_table.txt};
    \addlegendentry[]{\dumpsession};
    
    \addplot [black,fill=RandomColor,x tick label style={xshift=-0.3cm}, postaction={
        pattern=grid
    }] table[x=notebook,y=en] {sections/data/restore_undo_time_table.txt};
    \addlegendentry[]{\elasticnotebook};
    
    \addplot
    [black,fill=GreedyColor,x tick label style={xshift=-0.3cm}, postaction={
        pattern=crosshatch dots
    }] table[x=notebook,y=criuinc] {sections/data/restore_undo_time_table.txt};
    \addlegendentry[]{\criuincremental};

    \addplot [black,fill=BlueColor,x tick label style={xshift=-0.3cm}, postaction={
        pattern=bricks
    }] table[x=notebook,y=netezza] {sections/data/restore_undo_time_table.txt};
    \addlegendentry[]{\detreplay};
    
    \addplot [black,fill=HeuristicColor,x tick label style={xshift=-0.3cm}, postaction={
        pattern=crosshatch
    }] table[x=notebook,y=ours] {sections/data/restore_undo_time_table.txt};
    \addlegendentry[]{\system \textbf{(Ours)}};

    \node[anchor=south west, red, rotate=90,
    font=\footnotesize\bfseries] 
    at (axis cs: 6.22, 0.005) 
    {Error};

    \node[anchor=south west, red, rotate=90,
    font=\footnotesize\bfseries] 
    at (axis cs: 5.82, 0.005) 
    {Error};

    \node[anchor=south west, red, rotate=90,
    font=\footnotesize\bfseries] 
    at (axis cs: 4.94, 0.005) 
    {Error};

\end{axis}
    
\end{tikzpicture}
\vspace{-7mm}
\caption{Undo operations in notebooks (undone cell in brackets)}
\vspace{-3mm}
\label{fig:experiment_checkout_undo}
\end{subfigure}
\begin{subfigure}[b]{\linewidth}
\begin{tikzpicture}

\pgfplotstableread[col sep=comma,]{
name
Cluster (11)
TPS (34)
Sklearn (32)
HW-LM (63)
StoreSales (25)
TorchGPU (23)
}\datatable

\begin{axis}[
    ybar,
    clip=false,
    xtick={1, 2, 3, 4, 5, 6},
    xticklabels from table={\datatable}{name},
                 x tick label style={anchor=center, yshift = 0ex, font=\scriptsize},
    xlabel style={yshift = 2.5ex, font=\footnotesize},
    ylabel style={yshift = -3.5ex, font=\scriptsize},
    xtick style ={draw=none},
    width=90mm,
    height=26mm,
    bar width=1.0mm,
    ymin=0.01,
    ymax=100,
    log origin = infty,
    ymode = log,
    axis y line*=none,
    axis x line*=none,
    ytick={0.01, 0.1, 1, 10, 100},
    yticklabels={0.01, 0.1, 1, 10, 100},
    xmin=0.5,
    xmax = 6.5,
    ymajorgrids,
    tick label style={font=\footnotesize},
    legend style={
        font=\footnotesize,
        /tikz/every even column/.append style={column sep=0.2cm},
        legend columns = 3,
        at={(-0.06,1.1)},
        inner ysep=0.5pt,
        anchor=south west,
    },
    ylabel={Checkout Time (s)},
    area legend
    ]

    \addplot [black,fill=NoOptColor,x tick label style={xshift=-0.3cm}, postaction={
        pattern=north west lines
    }] table[x=notebook,y=criu] {sections/data/restore_branch_time_table.txt};

    \addplot [black,fill=LRUColor,x tick label style={xshift=-0.3cm}, postaction={
        pattern=horizontal lines
    }] table[x=notebook,y=dump] {sections/data/restore_branch_time_table.txt};
    
    \addplot [black,fill=RandomColor,x tick label style={xshift=-0.3cm}, postaction={
        pattern=grid
    }] table[x=notebook,y=en] {sections/data/restore_branch_time_table.txt};
    
    \addplot
    [black,fill=GreedyColor,x tick label style={xshift=-0.3cm}, postaction={
        pattern=crosshatch dots
    }] table[x=notebook,y=criuinc] {sections/data/restore_branch_time_table.txt};

    \addplot [black,fill=BlueColor,x tick label style={xshift=-0.3cm}, postaction={
        pattern=bricks
    }] table[x=notebook,y=netezza] {sections/data/restore_branch_time_table.txt};

    \addplot [black,fill=HeuristicColor,x tick label style={xshift=-0.3cm}, postaction={
        pattern=crosshatch
    }] table[x=notebook,y=ours] {sections/data/restore_branch_time_table.txt};

    \node[anchor=south west, red, rotate=90,
    font=\footnotesize\bfseries] 
    at (axis cs: 6.22, 0.005) 
    {Error};

    \node[anchor=south west, red, rotate=90,
    font=\footnotesize\bfseries] 
    at (axis cs: 5.82, 0.005) 
    {Error};

        \node[anchor=south west,
    font=\footnotesize\bfseries] 
    at (axis cs: 1.22, 20) 
    {1050s};
    

\end{axis}
    
\end{tikzpicture}
\vspace{-7mm}
\caption{Branch operation in notebooks (branch cell in brackets)}
\label{fig:experiment_checkout_branch}
\end{subfigure}

\caption{
Methods' checkout time for undoing executions (top) and switching to a branched states (bottom). \systembf's checkout is up to 8.18$\times$ and 4.18$\times$ faster, respectively, than the next best alternative; notably, the former is \textit{sub-second}.
}
\end{figure}
\subfile{plots/experiment_checkout_branch}
\subsubsection{Fast Path Exploration}
\label{sec:experiment_checkout_branch}

For each notebook, we (1) run the notebook end-to-end, (2) checkout to the state before any models are trained, (3) rerun to the end of the notebook (thus creating a second branch), and measure the time taken to switch back to the first branch containing different models and plots.

We report the results in \cref{fig:experiment_checkout_branch}. Similar to \cref{sec:experiment_checkout_undo}, \system performs sub-second branch switching on 4/6 notebooks by updating (only) models and plots differing between branches (i.e., not the input dataframes) and does so up to 4.18$\times$ faster than the next best alternative (\textit{StoreSales}).
While there is considerable divergence between branches in the \textit{StoreSales} test case (i.e., new auxiliary dataframes are created along ML models and plots), \system still performs branch switching at a fast 1.73 seconds, which is 4.18$\times$ faster than the next method (\dumpsession). While \detreplay can potentially be faster than \system (\textit{TorchGPU}) by replaying cells that allow it to bypass expensive data loading, it can also cause unacceptable checkout times (\textbf{1050s}, \textit{Cluster}, from replaying an entire deterministic model fitting sequence); hence, cost-based optimization is required for \detreplay to function, which we leave to future work.
\begin{table}[t]

\caption{\systemnosf's delta tracking time vs. baselines. \systemnosf tracks the per-execution delta up to 11.42$\times$ faster than the next best method and is only up to 2.03\% of notebook runtime.}
\footnotesize
\addtolength{\tabcolsep}{-3pt} 
\midsepremove
\begin{tabular}{|l|c|c|c|}
\toprule
 & \multicolumn{3}{c|}{Tracking overhead (s) (\% of notebook runtime)}\\
  \midrule
\textbf{Notebook} & \ipyflow & \systembad & \system \textbf{(Ours)}\\
 \midrule
Cluster &  5.091s (0.299\%) & 0.231s (0.014\%) & \textbf{0.094s (0.005\%)} \\
\midrule
TPS &  15.49s (10.06\%) & 4.618s (2.999\%) & \textbf{1.195s (0.776\%)} \\
\midrule
Sklearn &  11.75s (2.293\%) & 136.2s (26.61\%) & \textbf{10.39s (2.030\%)} \\
\midrule
HW-LM &  30.07s (231.3\%) & 2.964s (2.280\%) & \textbf{0.259s (1.992\%)} \\
\midrule
StoreSales &  \textbf{FAIL on cell 27} & 72.58s (10.91\%)& \textbf{11.24s (1.69\%)} \\
\midrule
QiSkit &  18.91s (41.12\%)& 2.235s (4.859\%)& \textbf{0.339s (0.737\%)} \\
\midrule
Ray & 12.90s (0.546\%)& 0.861s (0.036\%)& \textbf{0.199s (0.008\%)} \\
\midrule
TorchGPU & 3.731s (0.521\%)& 4.509s (0.630\%)& \textbf{0.578s (0.081\%)} \\
\bottomrule
\end{tabular}
\midsepdefault
\addtolength{\tabcolsep}{3pt}
\label{tbl:experiment_fast_capture}
\end{table}
\begin{figure}[t]\captionsetup[subfigure]{font=footnotesize}
\pgfplotsset{scaled y ticks=false}
\centering
\begin{subfigure}[b]{\linewidth}
\begin{tikzpicture}

\begin{axis}[
    xtick=data,
    width=85mm,
    height=26mm,
    ymin=0.0001,
    ymax=1000,
    log origin = infty,
    ymode = log,
    axis y line*=none,
    axis x line*=none,
    ytick={0.0001, 0.01, 1, 100},
    yticklabels={0.0001, 0.01, 1, 100},
    xlabel=Cell execution number,
    xlabel style={yshift = 2.5ex},
    ylabel style={yshift=-2ex},
    xmin = 1,
    xmax = 49,
    xtick = {1, 10, 20, 30, 40},
    xticklabels = {1, 10, 20, 30, 40},
    tick label style={font=\footnotesize},
    legend style={
        at={(-0.1,1.1)},anchor=south west,column sep=2pt,
        draw=black,fill=white,
        /tikz/every even column/.append style={column sep=5pt},
        inner ysep=0.5pt,
        font=\footnotesize
    },
    legend cell align={left},
    legend columns=5,
    label style={font=\footnotesize},
    ylabel={$\times$ of runtime},
    ymajorgrids,
    every axis plot/.append style={thick}
]

\draw[gray, thick, opacity=0.3] (axis cs: 22,0.0001) -- (axis cs: 22,1000);
\draw[gray, thick, opacity=0.3] (axis cs: 35,0.0001) -- (axis cs: 35,1000);
\draw[gray, thick, opacity=0.3] (axis cs: 43,0.0001) -- (axis cs: 43,1000);
\draw[gray, thick, opacity=0.3] (axis cs: 47,0.0001) -- (axis cs: 47,1000);

\addplot[GreenColor, mark = x, mark size=0.75pt, opacity = 0.7, densely dashed]
table[x=cellnum,y=ipyflow] {sections/testdata/percentage_tps_ipyflow.txt};
\addplot[RandomColor, mark = o, mark size=0.75pt, densely dotted, opacity = 0.7]
table[x=cellnum,y=en] {sections/testdata/percentage_tps_en.txt};
\addplot[HeuristicColor, mark = *, mark size=0.75pt, opacity = 0.7]
table[x=cellnum,y=ours] {sections/testdata/percentage_tps_kishu.txt};
\addlegendentry{\ipyflow}
\addlegendentry{\systembad}
\addlegendentry{\system \textbf{(Ours)}}



\end{axis}
\end{tikzpicture}
\vspace{-3mm}
\caption{TPS}
\vspace{-3mm}
\end{subfigure}
\hfill
\begin{subfigure}[b]{\linewidth}
\begin{tikzpicture}

\begin{axis}[
    xtick=data,
    width=85mm,
    height=26mm,
    ymin=0.0001,
    ymax=10000,
    log origin = infty,
    ymode = log,
    axis y line*=none,
    axis x line*=none,
    ytick={0.0001, 0.01, 1, 100, 10000},
    yticklabels={0.0001, 0.01, 1, 100, 10000},
    xlabel=Cell execution number,
    xlabel style={yshift = 2.5ex},
    ylabel style={yshift=-2ex},
    xmin = 1,
    xmax = 44,
    xtick = {1, 10, 20, 30, 40},
    xticklabels = {1, 10, 20, 30, 40},
    tick label style={font=\footnotesize},
    legend style={
        at={(-0.2,1.1)},anchor=south west,column sep=2pt,
        draw=black,fill=white,
        /tikz/every even column/.append style={column sep=5pt},
        font=\footnotesize
    },
    legend cell align={left},
    legend columns=5,
    label style={font=\footnotesize},
    ylabel={$\times$ of runtime},
    ymajorgrids,
    every axis plot/.append style={thick}
]
\draw[gray, thick, opacity=0.3] (axis cs: 5,0.0001) -- (axis cs: 5,10000);
\draw[gray, thick, opacity=0.3] (axis cs: 26,0.0001) -- (axis cs: 26,10000);
\addplot[GreenColor, mark = x, mark size=0.75pt, opacity = 0.7, densely dashed]
table[x=cellnum,y=ipyflow] {sections/testdata/percentage_sklearn_ipyflow.txt};
\addplot[RandomColor, mark = o, mark size=0.75pt, densely dotted, opacity = 0.7]
table[x=cellnum,y=en] {sections/testdata/percentage_sklearn_en.txt};
\addplot[HeuristicColor, mark = *, mark size=0.75pt, opacity = 0.7]
table[x=cellnum,y=ours] {sections/testdata/percentage_sklearn_kishu.txt};


\end{axis}
\end{tikzpicture}
\vspace{-3mm}
\caption{Sklearn}
\vspace{-3mm}
\end{subfigure}
\hfill
\hspace{-0.2mm}
\begin{subfigure}[b]{\linewidth}
\begin{tikzpicture}

\begin{axis}[
    xtick=data,
    width=85mm,
    height=26mm,
    ymin=0.001,
    ymax=10000,
    log origin = infty,
    ymode = log,
    axis y line*=none,
    axis x line*=none,
    ytick={0.0001, 0.01, 1, 100, 10000},
    yticklabels={0.0001, 0.01, 1, 100, 10000},
    xlabel=Cell execution number,
    xlabel style={yshift = 2.5ex},
    ylabel style={yshift=-2ex},
    xmin = 1,
    xmax = 81,
    xtick = {1, 20,40,60,80},
    xticklabels = {1, 20,40,60,80},
    tick label style={font=\footnotesize},
    legend style={
        at={(-0.2,1.1)},anchor=south west,column sep=2pt,
        draw=black,fill=white,
        /tikz/every even column/.append style={column sep=5pt},
        font=\footnotesize
    },
    legend cell align={left},
    legend columns=5,
    label style={font=\footnotesize},
    ylabel={$\times$ of runtime},
    ymajorgrids,
    every axis plot/.append style={thick}
]

\addplot[GreenColor, mark = x, mark size=0.75pt, opacity = 0.7, densely dashed]
table[x=cellnum,y=ipyflow] {sections/testdata/percentage_04_ipyflow.txt};
\addplot[RandomColor, mark = o, mark size=0.75pt, densely dotted, opacity = 0.7]
table[x=cellnum,y=en] {sections/testdata/percentage_04_en.txt};
\addplot[HeuristicColor, mark = *, mark size=0.75pt, opacity = 0.7]
table[x=cellnum,y=ours] {sections/testdata/percentage_04_kishu.txt};



\end{axis}
\end{tikzpicture}
\vspace{-3mm}
\caption{HW-LM}
\vspace{-2mm}
\end{subfigure}

\caption{Methods' per-cell tracking overhead as $\times$ of cell runtime. Gray vertical lines indicate long-running cells (>10s). \system outperforms baselines by efficiently detecting delta of long-running cells and identifying candidate updates (\cref{sec:checkpoint_fast})
}
\label{fig:experiment_fast_capture_percentage}
\end{figure}
\subsection{Fast Delta Detection}
\label{sec:experiment_fast_capture}

This section investigates \system's Co-variable granularity state tracking overhead by comparing the time taken by \system to track per-cell execution state delta with other tracking methods.

\paragraph{Cumulative Tracking Overhead (\cref{tbl:experiment_fast_capture})} \system is consistently fastest at detecting state delta and is (1) up to 11.42$\times$ faster than the best out of \ipyflow and \systembad (\textit{HW-LM}), and (2) only up to maximum of 2.03\% of notebook runtime (\textit{Sklearn}).
\paragraph{Per-Cell Tracking Overhead (\cref{fig:experiment_fast_capture_percentage})} 
We investigate per-cell execution tracking overhead of methods on selected notebooks. \system efficiently handles long-running cells (>10s, gray vertical lines) than \ipyflow: these cells often contain complex control flows (e.g., looped \texttt{if} statement, \textit{Sklearn} cells 5 and 26\footnote{StoreSales cell 27 contains complex control flows; \ipyflow hangs indefinitely.}) and/or call complex functions (e.g., model fitting, all 4 long cells in \textit{TPS}). As hypothesized in \cref{sec:motivation_delta}, IPyFlow incurs significant overhead on these cells (e.g., 0.3$\times$ on the 17s TPS cell 35) which \system circumvents by performing live analysis only between cell executions (0.3$\times \rightarrow$ 0.001$\times$).

Compared to \systembad, \system identifies and only checks possibly updated Co-variables (\cref{sec:checkpoint_fast}): this exploits the incremental nature of cell executions (\cref{sec:motivation_workload}) and is \textit{necessary}---in \textit{Sklearn}, \systembad's detection overhead grows significantly as more objects are introduced into the kernel (up to 4936$\times$, cell 42), while \system's approach bounds the overhead (4936$\times \rightarrow$ 0.84$\times$).

Notably, there exists further optimization opportunities for \system such as for (1) cells updating Co-variables with nested \vargraphs (e.g. list of strings \texttt{text\_neg}, \textit{Sklearn} cell 41) and (2) read-only printing cells (e.g., \texttt{y\_train[:10]}, \textit{HW-LM} cell 67). \system incurs significant overhead on these cases (260$\times$ and 1.06$\times$ respectively; while they are of low absolute value due to the short cell runtimes (547ms/2ms and 2ms/2ms respectively), this indicates need for more efficient delta detection methods such as list hashing and rule-based detection, respectively (\cref{sec:implementation_consideration}), which we leave for future exploration.





\subsection{Workload Study}
\begin{table}[t]

\caption{Variable vs. Co-variable count in notebooks.}
\footnotesize
\addtolength{\tabcolsep}{-3.5pt} 
\midsepremove
\begin{tabular}{lrr|lrr|lrr}
\toprule
\textbf{Notebook} & \textbf{var\#} & \textbf{Co-var\#} & \textbf{Notebook} & \textbf{var\#} & \textbf{Co-var\#}   & \textbf{Notebook} & \textbf{var\#} & \textbf{Co-var\#} \\
 \midrule
Cluster & 28 & 26 & TPS & 47 & 47 & Sklearn & 43 & 42 \\
HW-LM & 172 & 170 & StoreSales & 27 & 27 & Qiskit & 51 & 41  \\
 Ray & 29 & 28 & TorchGPU & 57 & 52 \\
\bottomrule
\end{tabular}
\midsepdefault
\addtolength{\tabcolsep}{3.5pt}
\label{tbl:experiment_covariable_count}
\end{table}
\begin{figure}[t]\captionsetup[subfigure]{font=footnotesize}
\pgfplotsset{scaled y ticks=false}
\centering
\begin{subfigure}[b]{0.48\linewidth}
\begin{tikzpicture}

\begin{axis}[
    xtick=data,
    width=45mm,
    height=26mm,
    ymin=0,
    ymax=4,
    axis y line*=none,
    axis x line*=none,
    xtick={1,2,3,4,5, 6},
    xticklabel style = {align=center},
    xticklabels = {1600, 800, 400, 200, 100, 50},
    ytick={0, 1, 2, 3, 4},
    yticklabels={0, 1, 2, 3, 4},
    xlabel=\% of updated data,
    xlabel style={yshift = 2.5ex,font=\footnotesize},
    ylabel style={yshift=-4ex,xshift=-0.5ex,font=\scriptsize},
    xmin = 0,
    xmax = 100,
    xtick = {0, 20, 40,60, 80, 100},
    xticklabels = {0, 20, 40,60, 80, 100},
    tick label style={font=\footnotesize},
    legend style={
        at={(-0.2,1.1)},anchor=south west,column sep=2pt,
        draw=black,fill=white,
        /tikz/every even column/.append style={column sep=5pt},
        inner ysep=0.5pt,
        font=\scriptsize,
    },
    legend cell align={left},
    legend columns=4,
    ylabel={Avg. Time (s)},
    ylabel style={yshift=-1.5ex},
    ymajorgrids,
    every axis plot/.append style={thick}
]

\addplot[GreenColor, mark = x, mark size=0.75pt, opacity = 0.7, densely dashed]
table[x=percent,y=dill, densely dashed] {sections/data/covariable_time.txt};
\addlegendentry{\dumpsession}
\addplot[RandomColor, mark = o, mark size=0.75pt, densely dotted, opacity = 0.7]
table[x=percent,y=criu, densely dashed] {sections/data/covariable_time.txt};
\addlegendentry{\criuincremental}
\addplot[HeuristicColor, mark = *, mark size=0.75pt, opacity = 0.7]
table[x=percent,y=ours] {sections/data/covariable_time.txt};
\addlegendentry{\system \textbf{(Ours)}}

\draw[black, very thick, opacity=0.5] (axis cs: 2.57,0) -- (axis cs: 2.57,4);


\end{axis}
\end{tikzpicture}
\vspace{-7mm}
\caption{Checkpoint time}
\vspace{-2mm}
\end{subfigure}
\begin{subfigure}[b]{0.48\linewidth}
\begin{tikzpicture}

\begin{axis}[
    xtick=data,
    width=45mm,
    height=26mm,
    ymin=0,
    ymax=750000000,
    axis y line*=none,
    axis x line*=none,
    xtick={1,2,3,4,5, 6},
    xticklabel style   = {align=center},
    xticklabels = {1600, 800, 400, 200, 100, 50},
    ytick={0, 250000000, 500000000, 750000000},
    yticklabels={0, 250, 500, 750},
    xlabel=\% of updated data,
    xlabel style={yshift = 2.5ex,font=\footnotesize},
    ylabel style={yshift=-4ex,xshift=-0.5ex,font=\scriptsize},
    xmin = 0,
    xmax = 100,
    xtick = {0, 20, 40,60, 80, 100},
    xticklabels = {0, 20, 40,60, 80, 100},
    tick label style={font=\footnotesize},
    legend style={
        at={(-0.2,1.1)},anchor=south west,column sep=2pt,
        draw=black,fill=white,
        /tikz/every even column/.append style={column sep=5pt},
        font=\scriptsize,
    },
    legend cell align={left},
    legend columns=4,
    ylabel={Total size (MB)},
    ylabel style={yshift=-1ex},
    ymajorgrids,
    every axis plot/.append style={thick}
]

\addplot[GreenColor, mark = x, mark size=0.75pt, opacity = 0.7, densely dashed]
table[x=percent,y=dill] {sections/data/covariable_size.txt};
\addplot[RandomColor, mark = o, mark size=0.75pt, densely dotted, opacity = 0.7]
table[x=percent,y=criu, densely dashed] {sections/data/covariable_size.txt};
\addplot[HeuristicColor, mark = *, mark size=0.75pt, opacity = 0.7]
table[x=percent,y=ours] {sections/data/covariable_size.txt};

\draw[black, very thick, opacity=0.5] (axis cs: 2.57,0) -- (axis cs: 2.57,750000000);


\end{axis}
\end{tikzpicture}
\vspace{-3mm}
\caption{Checkpoint size}
\vspace{-2mm}
\end{subfigure}
\hfill
\begin{subfigure}[b]{0.48\linewidth}
\begin{tikzpicture}

\begin{axis}[
    xtick=data,
    width=45mm,
    height=26mm,
    ymin=0,
    ymax=3.0,
    axis y line*=none,
    axis x line*=none,
    xtick={1,2,3,4,5, 6},
    xticklabel style   = {align=center},
    xticklabels = {1600, 800, 400, 200, 100, 50},
    ytick={0, 0.5, 1, 1.5, 2, 2.5, 3.0},
    yticklabels={0, 0.5, 1, 1.5, 2, 2.5, 3.0},
    xlabel=\% of updated data,
    xlabel style={yshift = 2.5ex,font=\footnotesize},
    ylabel style={yshift=-4ex,xshift=-0.5ex,font=\scriptsize},
    xmin = 0,
    xmax = 100,
    xtick = {0, 20, 40,60, 80, 100},
    xticklabels = {0, 20, 40,60, 80, 100},
    ylabel style={yshift=-1ex},
    tick label style={font=\footnotesize},
    legend style={
        at={(-0.2,1.1)},anchor=south west,column sep=2pt,
        draw=black,fill=white,
        /tikz/every even column/.append style={column sep=5pt},
        font=\scriptsize,
    },
    legend cell align={left},
    legend columns=4,
    ylabel={Avg. Time(s)},
    ymajorgrids,
    every axis plot/.append style={thick}
]

\addplot[GreenColor, mark = x, mark size=0.75pt, opacity = 0.7, densely dashed]
table[x=percent,y=dill] {sections/data/covariable_restore.txt};
\addplot[RandomColor, mark = o, mark size=0.75pt, densely dotted, opacity = 0.7]
table[x=percent,y=criu, densely dashed] {sections/data/covariable_restore.txt};
\addplot[HeuristicColor, mark = *, mark size=0.75pt, opacity = 0.7]
table[x=percent,y=ours] {sections/data/covariable_restore.txt};

\draw[black, very thick, opacity=0.5] (axis cs: 2.57,0) -- (axis cs: 2.57,3.0);



\end{axis}
\end{tikzpicture}
\vspace{-3mm}
\caption{Checkout time}
\end{subfigure}
\begin{subfigure}[b]{0.48\linewidth}
\begin{tikzpicture}

\begin{axis}[
    xtick=data,
    width=45mm,
    height=26mm,
    ymin=0,
    ymax=0.1,
    axis y line*=none,
    axis x line*=none,
    xtick={1,2,3,4,5, 6},
    xticklabel style   = {align=center},
    xticklabels = {1600, 800, 400, 200, 100, 50},
    ytick={0, 0.02, 0.04, 0.06, 0.08, 0.1},
    yticklabels={0, 20, 40, 60, 80, 100},
    xlabel=\% of updated data,
    xlabel style={yshift = 2.5ex,font=\footnotesize},
    ylabel style={yshift=-4ex,xshift=-0.5ex,font=\scriptsize},
    xmin = 0,
    xmax = 100,
    xtick = {0, 20, 40,60, 80, 100},
    xticklabels = {0, 20, 40,60, 80, 100},
    ylabel style={yshift=-1ex},
    tick label style={font=\footnotesize},
    legend style={
        at={(-0.2,1.1)},anchor=south west,column sep=2pt,
        draw=black,fill=white,
        /tikz/every even column/.append style={column sep=5pt},
        font=\scriptsize,
    },
    legend cell align={left},
    legend columns=4,
    ylabel={Avg. Time(ms)},
    ymajorgrids,
    every axis plot/.append style={thick}
]

\addplot[HeuristicColor, mark = *, mark size=0.75pt, opacity = 0.7]
table[x=percent,y=ours] {sections/data/covariable_profiling.txt};

\draw[black, very thick, opacity=0.5] (axis cs: 2.57,0) -- (axis cs: 2.57,0.1);

\end{axis}
\end{tikzpicture}
\vspace{-3mm}
\caption{Tracking time}
\end{subfigure}

\caption{Checkpoint/checkout efficiency vs. \% of data in updated list Co-variable. \system performs best when each Co-variable contains few data in the state, which is typical of real-world notebooks (\cref{tbl:experiment_covariable_count}, vertical line marks average).
}
\label{fig:experiment_covariable}
\end{figure}
This section studies \system's performance versus parameter sweeps on the degree of shared referencing between variables (i.e., Co-Variable size) (\cref{sec:workload_study_covariable}) and number of cell executions (\cref{sec:workload_study_long}).

\subsubsection{Performance vs. Shared Referencing}
\label{sec:workload_study_covariable}
We insert ten 64MB numpy arrays into a list (\textit{\% of state data in a Co-Variable}). We evaluate \system's checkpoint/checkout costs on cells modifying only \textit{one} array in the list/Co-Variable. \dumpsession's and \criuincremental's performances are provided as comparison.

We report results in \cref{fig:experiment_covariable}. As \system detects updates and checkpoints at the Co-variable granularity, it checks for updates and then checkpoints all arrays in the list after each test cell. \system performs best when the list contains a low \% of data in the state, as it can (1) limit the scope of update checking and (2) time-travel by saving/loading a small amount of state data. \system's performance drops as the list Co-variable bundles the changed array with more unchanged data and is equivalent to \dumpsession \footnote{Different serialization libraries cause different checkpoint/checkout time cost; there is also overhead with managing data with blob storage vs. writing directly to file (\cref{sec:implementation_details}).} when all 10 arrays (i.e., all data in the state) form one large Co-Variable: \system has to (1) check the whole state for updates and (2) save/update all data for checkpoint/checkout, while \criuincremental can still checkpoint only the one changed array in the (640MB) Co-variable.

However, as our evaluation workloads suggest (\cref{tbl:experiment_covariable_count}), 
states typically consist of a large number of small Co-Variables 
    (each containing 2.57\% of state data on average, black vertical lines in \cref{fig:experiment_covariable}).
For these typical cases, our Co-Variable-based approach
    significantly outperforms other baselines (\cref{sec:experiment_cheap_checkpoint}, \cref{sec:experiment_fast_checkpoint}).

\subsubsection{Scalability to Long Notebook Sessions}
\label{sec:workload_study_long}
\begin{figure}[t]\captionsetup[subfigure]{font=footnotesize}
\pgfplotsset{scaled y ticks=false}
\centering
\begin{subfigure}[b]{0.48\linewidth}
\begin{tikzpicture}

\begin{axis}[
    xtick=data,
    width=45mm,
    height=26mm,
    ymin=0,
    ymax=10000000,
    axis y line*=none,
    axis x line*=none,
    xtick={1,2,3,4,5, 6},
    xticklabel style   = {align=center,font=\scriptsize},
    xticklabels = {1600, 800, 400, 200, 100, 50},
    ytick={0, 2000000, 4000000, 6000000, 8000000, 10000000},
    yticklabels={0,2,4,6,8,10},
    xlabel=Cell executions,
    xlabel style={yshift = 2.5ex},
    ylabel style={yshift=-4ex},
    xmin = 0,
    xmax = 1000,
    xtick = {0, 250, 500, 750, 1000},
    xticklabels = {0, 250, 500, 750, 1000},
    tick label style={font=\footnotesize},
    legend style={
        at={(-0.2,1.1)},anchor=south west,column sep=2pt,
        draw=black,fill=white,
        /tikz/every even column/.append style={column sep=5pt},
        inner ysep=0.5pt,
        font=\scriptsize,
    },
    legend cell align={left},
    legend columns=4,
    label style={font=\footnotesize},
    ylabel={Size (MB)},
    ylabel style={yshift=-1.5ex},
    ymajorgrids,
    every axis plot/.append style={thick}
]

\addplot[GreenColor, mark = x, mark size=0.75pt, opacity = 0.7, densely dashed]
table[x=cell,y=hwlm, densely dashed] {sections/data/checkpoint_graph_size.txt};
\addlegendentry{HW-LM}
\addplot[HeuristicColor, mark = *, mark size=0.75pt, opacity = 0.7]
table[x=cell,y=qiskit] {sections/data/checkpoint_graph_size.txt};
\addlegendentry{Qiskit}
\label{fig:experiment_scalability_a}


\end{axis}
\end{tikzpicture}
\vspace{-2.5mm}
\caption{Size of Checkpoint Graph}
\end{subfigure}
\begin{subfigure}[b]{0.48\linewidth}
\begin{tikzpicture}

\begin{axis}[
    xtick=data,
    width=45mm,
    height=26mm,
    ymin=0,
    ymax=0.1,
    axis y line*=none,
    axis x line*=none,
    xtick={1,2,3,4,5, 6},
    xticklabel style   = {align=center,font=\scriptsize},
    xticklabels = {1600, 800, 400, 200, 100, 50},
    ytick={0, 0.02, 0.04, 0.06, 0.08, 0.1},
    yticklabels={0, 20, 40, 60, 80, 100},
    xlabel=No.~cells in target state,
    xlabel style={yshift = 2.5ex},
    ylabel style={yshift=-4ex},
    xmin = 50,
    xmax = 950,
    xtick = {50, 250, 500, 750, 950},
    xticklabels = {50, 250, 500, 750, 950},
    ylabel style={yshift=-1ex},
    tick label style={font=\footnotesize},
    legend style={
        at={(-0.2,1.1)},anchor=south west,column sep=2pt,
        draw=black,fill=white,
        /tikz/every even column/.append style={column sep=5pt},
        font=\scriptsize,
    },
    legend cell align={left},
    legend columns=4,
    label style={font=\footnotesize},
    ylabel={Find time (ms)},
    ymajorgrids,
    every axis plot/.append style={thick}
]

\addplot[GreenColor, mark = x, mark size=0.75pt, opacity = 0.7, densely dashed]
table[x=cell,y=hwlm, densely dashed] {sections/data/undo_far_time.txt};
\addplot[HeuristicColor, mark = *, mark size=0.75pt, opacity = 0.7]
table[x=cell,y=qiskit] {sections/data/undo_far_time.txt};
\label{fig:experiment_scalability_c}


\end{axis}
\end{tikzpicture}
\vspace{-2.5mm}
\caption{Find diff. for undo time @ 1K CEs}
\end{subfigure}

\caption{\system's scalability vs. cell executions. Checkpoint Graph size increases linearly; state difference computation time is linear vs. total cell count in current and target states.
}
\label{fig:experiment_scalability}
\end{figure}
We choose 2 Data Visualization notebooks (\textit{HW-LM} and \textit{Qiskit}, and randomly re-execute up to 1000 cells.\footnote{This is the length of the longest observed notebook on Kaggle~\cite{ai4code}, and $\sim$10 times the 97th percentile cell execution count of a notebook workload~\cite{rule2018exploration}.} We measure the (1) Checkpoint Graph size and (2) time to compute state difference for undoing 0-1000 cells (i.e., to a prior state) from the state after the 1000th cell execution.

We report results in \cref{fig:experiment_scalability}. The size of \system's Checkpoint Graph scales linearly with the number of executed cells and is up to only 9MB at 1000 cells (\cref{fig:experiment_scalability_a}, \textit{HW-LM}). \system's time to compute state difference between current and target states scales linearly with the \textit{sum of cell count in the two states}, up to 81ms for any checkout operation in a 1000 cell execution-long session.\footnote{This is because \system traverses the Checkpoint Graph to find the lowest common ancestor state on checkout with an off-the-shelf algorithm~\cite{treelca}.} These overheads are negligible on a more typical notebook---only 133KB and 5.8ms, respectively, on our longest experiment notebook (\textit{Qiskit}, 85 cells).

\section{Related Work}
\label{sec:related}

This section covers related work in PITR/checkpointing in various database applications (\cref{sec:related_work_db}) and related notebook systems and their employed techniques for lineage tracing (\cref{sec:related_work_notebook}).
, and other applicable tools for saving/loading data in notebook states (\cref{sec:related_work_serde})

\subsection{PITR and Incremental Checkpointing}
\label{sec:related_work_db}

\paragraph{PITR in Relational Databases}

In many DBMSs, 
    mechanisms like ARIES~\cite{mohan1992aries, mohan1988recovery} and 
        its variants~\cite{arulraj2016write,zheng2014fast, soroush2013time, schule2019versioning, gray1981recovery, 
        salem1989checkpointing, verhofstad1978recovery, morrey2003peabody}
    achieve durability and Point-in-time-Recovery (PITR) by 
combining (incremental) checkpointing and physical/operation logging~\cite{mysql, postgresql, soroush2013time, schule2019versioning, xu2017orpheusdb, antonopoulos2019constant, verbitski2017amazon, zheng2014fast}.
Persisting dirty objects (e.g., rows in relations) 
    can enhance recovery efficiency by reducing the number of log entries to replay.
ARIES~\cite{mohan1992aries} identifies dirty objects at the page level
    by recording a
    RecLSN, i.e.,
        the earliest modification time, in the dirty page table.
This is possible because all the pages 
    are controlled by the transaction system's buffer manager.
\system shares similarities:
    updates are periodically flushed. 
However, 
    the core question is \emph{how to define those updates} in computational notebooks.
There are no buffer managers;
    moreover, variables/objects are dependent with inter-object references (i.e., memory pointers).
Simply storing/restoring some variables independently thus can invalidate the state.
\system addresses this with Co-Variables,
    atomic units of persistence
        that can correctly preserve inter-object dependencies.
\system can still benefit from other PITR optimizations, 
    e.g., non-blocking checkpointing/restoration~\cite{mohan1992aries} and 
cost-based deterministic replay~\cite{netezzanondet} (\cref{sec:experiment_checkout_branch}), which we leave as future work.


\paragraph{PITR in Blob Storages}
Blob storages have incorporated PITR 
for recovering its state to earlier points~\cite{netezzanondet, postgresqlpitr, azurechangefeed, azurepitr, aurorapitrpdf, aurorapitr, sqlitepitr}.
These systems typically log timestamped updates (e.g., Azure's change feeds~\cite{azurechangefeed} and SQLite's binary log~\cite{sqlitepitr}), which can be used to return the storage state to a previous snapshot.
Computational notebooks require different approaches to determining state deltas,
    and \system addresses the unique challenge through Co-Variables and related techniques, which is the core novelty of this work.


\paragraph{Incremental and Differential Checkpointing in HPC}
HPC and stream processing systems implement efficient incremental checkpointing~\cite{lin2020incremental, keller2019application, bronevetsky2008compiler, apacheflink}. 
These works focus on high-frequency (e.g., >1M/second) updates to relatively simple data structures (e.g., KV-stores~\cite{lin2020incremental, apacheflink} and datasets~\cite{keller2019application, bronevetsky2008compiler}).
For enhanced efficiency,
    log truncation~\cite{lin2020incremental} and copying~\cite{keller2019application} 
    are employed
    to ensure the bounded size and robustness of incremental logs.
Computational notebooks require different delta detection methods 
as notebooks involve more complex objects, e.g., graphs, tensors, dataframes
    (\cref{sec:motivation_workload}).
While orthogonal, the existing optimization techniques for HPC may offer
    significant performance benefits
        if the core techniques for \system are expanded
            to handle a significantly larger number of executions (e.g., scripts)
        compared to interactive notebooks.

\subsection{Notebook Systems and Techniques}
\label{sec:related_work_notebook}
\paragraph{Systems for Speeding up Data Exploration}

There are a variety of works for enhancing data exploration efficiency in notebook-based systems~\cite{wu2020b2, koop2017dataflow, shankar2022bolt, lee2021lux, li2023edassistant, bauerle2022symphony, guo2012burrito, brachmann2019data, chockchowwat2023transactional, kishuboard}.
Reactive execution engines~\cite{wu2020b2, koop2017dataflow, shankar2022bolt} track cell reruns and rerun their dependent cells reactively to enforce consistent cell outputs.
Notebook recommender systems~\cite{lee2021lux, li2023edassistant} compute next-cell recommendations based on the current workflow.
Symphony~\cite{bauerle2022symphony} and B2~\cite{wu2020b2} enable point-and-click interactions with ML models and dataframes, respectively, by translation to equivalent code operations.
Burrito~\cite{guo2012burrito} and Vizier~\cite{brachmann2019data} construct and visualize a history graph for versioning multi-language (Python, R, SQL) data science code.
Diff-in-the-loop~\cite{wang2022diff} enables graphical comparisons of dataframes.
\system facilitates data exploration through efficient time-traveling, 
    and thus, is orthogonal and complementary to these works.
\paragraph{Notebook State Versioning}
Notebook state versioning has been explored by ForkIt~\cite{weinman2021fork}, which performs backtracking and branching to/from states by saving entire states with pickle~\cite{pickle}.
Our work versions notebook states in a much more efficient and robust manner compared to Forkit's approach (which is equivalent to our \dumpsession baseline in \cref{sec:experiments}) through efficient delta detection and incremental checkpointing and checkout.


\paragraph{Lineage Tracing in Notebooks}
Lineage tracing aims to capture code dependencies, i.e., accessed and updated data of cells, and has been widely used in notebook systems for a variety of downstream tasks (e.g., reactive execution)~\cite{manne2022chex, macke2020fine, ahmad2022reproducible, lielasticnotebook, cunha2021context, wu2020b2, koop2017dataflow, shankar2022bolt, lee2021lux, li2023edassistant, brachmann2019data, deo2022runtime}.
Tracing methods can be divided into (1) static code analysis using methods like AST decomposition~\cite{ast}, and (2) live code instrumentation resolving variable/data references at runtime.
They are relatively (1) cheap but conservative, and (2) accurate but expensive, respectively.
Lineage tracers often combine static and live analysis, 
to mitigate conservative assumptions of static analysis~\cite{lielasticnotebook} 
with dynamically computed 'ground truths'~\cite{macke2020fine, shankar2022bolt,deo2022runtime}.
Unfortunately, to the best of our knowledge, all of these works,
    except for ElasticNotebook~\cite{lielasticnotebook},
    detect modifications at the variable level, 
        incorrectly disregard shared references between variables.
\system's innovations in lineage tracing is 
    efficient live analysis 
    through (1) modeling the state at a coarser Co-variable granularity 
    and 
    (2) quickly pruning update candidates. 
This allows \system to achieve low false positives (\cref{sec:experiments_robust})
    with low overhead (\cref{sec:experiment_fast_capture}). 


\subsection{Checkpointing Notebook Objects}
\label{sec:related_work_serde}
\paragraph{Data Serialization for Checkpoint/Restore}

Data in IPython-based (e.g., Jupyter) notebook session states can be saved with serialization libraries~\cite{pickle, dill, marshal, pythonjson, bson, msgpack, serpy, cloudpickle}, on which a variety of existing checkpointing tools are built: On-disk KV-stores save individual variables~\cite{jupyterstore, pythonshove, redisshelve, pythonshelve, zodb, pythonchest}, DumpSession~\cite{dumpsession} saves the session state in bulk, and ElasticNotebook~\cite{lielasticnotebook} combines data storage/loading with cell replay for optimized session replication.
These works do not checkpoint nor restore incrementally, or have limitations/require significant user effort: 
Dill and ElasticNotebook's checkpoint files must be entirely loaded for restoration; while KV stores can store and load parts of a state, delta detection and shared reference preservation must be manually handled. 
In comparison, \system can perform low-overhead incremental checkpointing and checkout while preserving shared references (i.e., correctness) and works with almost all data science libraries (\cref{sec:experiments_robust}).

\paragraph{Memory Snapshotting}

There exist OS-level checkpointing tools that can incrementally checkpoint a process for later restoration~\cite{criu, ansel2009dmtcp, chen1997clip, plank1994libckpt, garg2018crum,jain2020crac, li2010frem}. These tools identify and store dirty memory pages, then piece together the process image for restoration~\cite{criu}: they often (1) cause large checkpoint sizes due to coarse page-level deltas~\cite{elnozahy2002survey}, (2) can only handle single processes~\cite{criulimitations}, and (3) can only restore from scratch: while we found a patent~\cite{neary2007method} and paper~\cite{ferreira2011libhashckpt} enabling incremental checkpointing for multiprocessing jobs and with sub-memory-page granularity, respectively, we could not locate working implementations.
In comparison, \system achieves lower checkpoint overheads via the logical Co-variable granularity deltas (\cref{sec:experiment_cheap_checkpoint}), checkpoint multiprocessing notebooks via application-level instructions (\cref{sec:experiments_robust}), and incrementally restores via state difference detection and using existing kernel data (\cref{sec:experiment_checkout}).
\section{Conclusion}

We have proposed \system,
a new computational notebook system
    that offers efficient and fault-tolerant time-traveling between notebook states.
    \system captures session state evolution at a novel Co-variable granularity for efficient incremental checkpointing of state deltas, which \system then uses to perform incremental checkout with minimal data loading.
    \system's contributions
    include (1) low-overhead state delta detection, (2) branch-based state versioning, and (3) generalizable time-traveling---preserving inter-variable dependencies and handling missing data with fallback recomputation. 
We have shown that \system is compatible with 146 data science object classes and reduces checkpoint storage size and checkout time by up to 4.55$\times$ and 8.18$\times$, respectively, 
on real-world notebooks.




\balance
\bibliographystyle{ACM-Reference-Format}
\bibliography{main}
\section{Appendix}
\label{sec:appendix}
\subsection{Notebook Categorization}
\label{appendix:percell}

\begin{figure}[t]
	\centering	\includegraphics[width=0.5\textwidth]{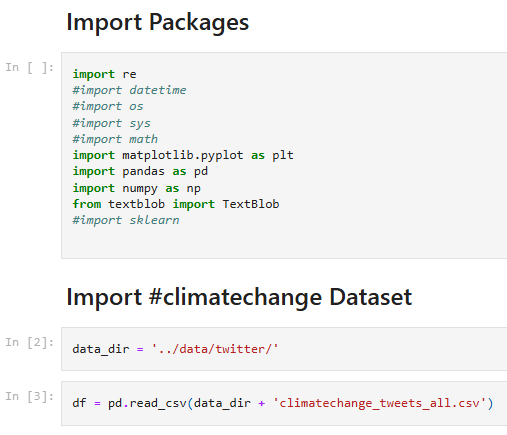}
	\caption{Hidden state in the in-progress \textit{Sklearn} notebook; it is uncertain (from the cell execution counts alone) what the first executed cell was.}
	\label{fig:appendix_debug}
\end{figure}

\begin{figure}[t]
	\centering	\includegraphics[width=0.5\textwidth]{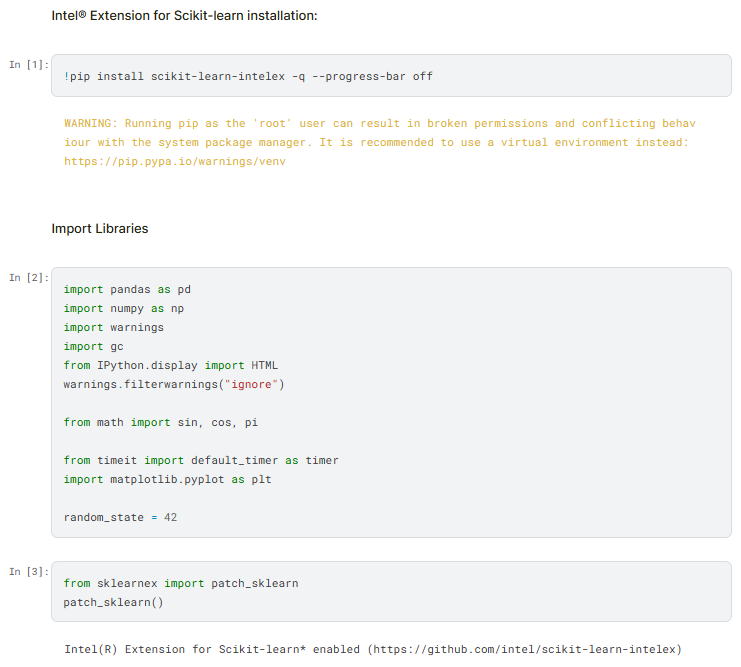}
	\caption{The final \textit{TPS} notebook with linear cell execution counts; also note the lack of commented-out 'draft' lines.}
	\label{fig:appendix_final}
\end{figure}

\paragraph{In-progress Versus Final Notebooks}
In our evaluation (\cref{sec:exp_setup}), we categorize in-progress vs. final notebooks (\cref{tbl:workload}) based on the existence of \textit{hidden states}~\cite{johnson2020benefits} (i.e., non-linear execution counts), a key characteristic of in-progress notebooks. These non-linear counts often occur from out-of-order and/or (re-)executions of the same cell performed for fast iteration~\cite{kery2018story} (e.g., debugging after a runtime error, adjusting prior imports~\cite{johnson2020benefits}). Hence, as observed in \cite{kery2018story}, a common step of notebook cleaning (i.e., from in-progress to final) is to tidy the code cells (e.g., via adding comments/deleting redundant code), then perform a clean run of all cells (e.g., via JupyterLab~\cite{jupyter}'s 'Run All' button) to verify that running the cells sequentially produces the desired final result. This is because notebook cleaning is often performed to ensure the reproducibility, which out-of-order executions/hidden states can potentially prevent~\cite{rule2018exploration}. For example, in \cref{fig:appendix_debug}, the in-progress notebook \textit{Sklearn} contains hidden states due to the absence of cell execution 1, while the final notebook \textit{TPS} (\cref{fig:appendix_final}) has sequentially-numbered cell executions. We summarize our categorization in \cref{tbl:appendix}.

\system works well for both in-progress and final notebooks (\cref{sec:experiments}); this is because, despite the distinct categorization, data science notebooks (whether final or not) share the key trait of containing many incremental cell executions (\cref{sec:motivation_use_case}), which we discuss shortly.

\begin{table}[t]

\caption{Notebooks by category and associated traits.}
\vspace{-3mm}
\footnotesize
\midsepremove
\begin{tabular}{llrr}
\toprule
\textbf{Notebook} & \textbf{Final} & \textbf{Hidden States} & \textbf{Out-of-order Cells} \\
 \midrule
Cluster\cite{bruteforce}& Yes & 0 & 0 \\
TPS\cite{tpsmar} & Yes & 0 & 0\\
HW-LM\cite{hwlm} & Yes & 0 & 0\\
StoreSales\cite{storesales}& Yes & 0 & 0\\
TorchGPU\cite{pytorchresnet} & Yes & 0 & 0\\ 
Sklearn\cite{sklearntweet}& No & 1 & 2 \\
Qiskit\cite{qiskit}& No & 91 & 1\\
Ray\cite{raynb} & No & 1 & 0 \\
\bottomrule
\end{tabular}
\midsepdefault
\vspace{-2mm}
\label{tbl:appendix}
\end{table}
\begin{figure}[t]
	\centering	\includegraphics[width=0.5\textwidth]{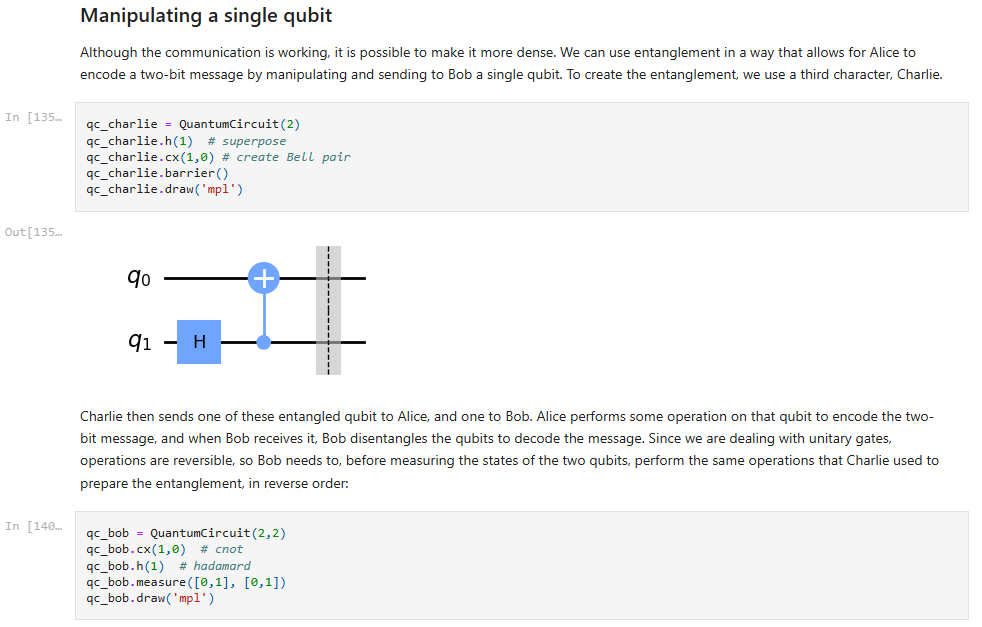}
	\caption{Likely re-execution of cell 140 in the in-progress \textit{Qiskit} notebook to adjust the plotting of qubit \textit{qc\_bob}.}
	\label{fig:appendix_reexecute}
\end{figure}

\begin{figure}[t]
	\centering	
    \includegraphics[width=0.5\textwidth]{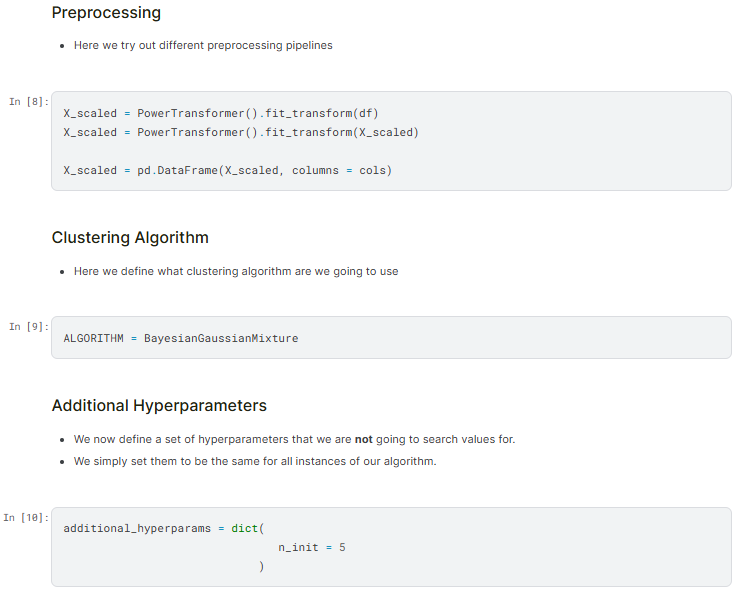}
	\caption{Granular yet incremental statements in the final \textit{Bruteforce} notebook. Each cell is a minimal data operation.}
	\label{fig:appendix_granular}
\end{figure}

\begin{figure}[t]
	\centering	
    \includegraphics[width=0.5\textwidth]{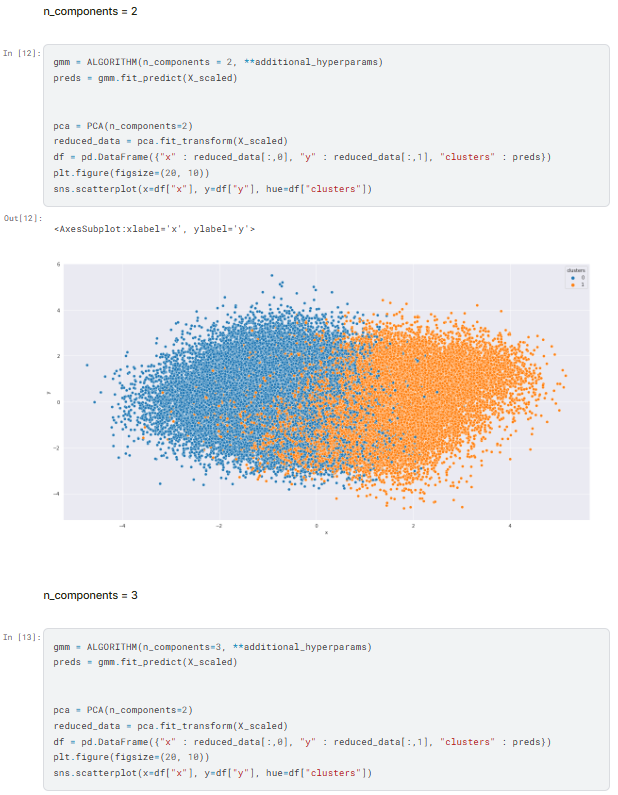}
	\caption{Granular model training in the \textit{Bruteforce} notebook. Despite their similarities, each cell trains only one model.}
	\label{fig:appendix_granular2}
\end{figure}

\paragraph{Incremental Cells in In-progress and Final Notebooks}
Both in-progress and final notebooks contain many incremental cell executions~\cite{kery2018story, rule2018exploration}, which are key to \system's capability to perform efficient time-traveling for notebook workloads. For example, incremental executions in in-progress notebooks can occur from repeated re-executions of the same cell with minor adjustments (e.g., axis scales) and buildups (e.g., adding a title)~\cite{kery2018story}, which \cite{eichmann2020idebench} observes to be particularly common for data visualization with the average plot being modified 7 times. This is seen in the in-progress notebook \textit{Qiskit} (\cref{fig:appendix_reexecute}), where we can infer that cell 140 for drawing \texttt{qc\_bob} has been sequentially executed 5 times\footnote{Unfortunately, the authors cannot be certain about their claim as the Qiskit notebook does not contain the full workload trace (i.e., cell executions 136-139). To the best of the authors' knowledge, there is no publically available notebook dataset that contains code for hidden cell (re-)executions; however, if availability allows, conducting a study on Kishu with one can be valuable future work.} before the desired end result was reached. The same incremental cell pattern is similarly present in final notebooks; while the notebooks are typically intended for linear execution hence re-executions being absent, users commonly organize their code in \textit{logical blocks}~\cite{kery2018story} where each cell is a \textit{minimal} standalone operation such as loading (only) one dataset and drawing (only) one plot, for ease of understanding and debugging, i.e., it is rare for (long) cells to contain multiple operations (e.g., data loading and model training in the same cell). This is observed in the final notebook \textit{Bruteforce} where each cell performs a granular task: in \cref{fig:appendix_granular}, the code for preprocessing, importing the clustering algorithm, and setting the hyperparameters are in different cells, with each accessing only 1-2 variables, while in \cref{fig:appendix_granular2}, each cell trains (only) one model with a specific set of hyperparameters (over)written into the same group of variables each time. Quantitatively,  \cref{fig:appendix_workload_characteristics} shows that cell executions in the final \textit{TPS} notebook are indeed incremental---each accessing only a part of the constantly growing session state (and additionally, balances of data creation and modification), which are also characteristics shown in the cell executions of the in-progress \textit{Sklearn} notebook (\cref{fig:background_workload_characteristics}).

\begin{figure}[t]
\pgfplotsset{scaled y ticks=false}
\centering
\begin{subfigure}[b]{0.48\linewidth}
\begin{tikzpicture}

\begin{axis}[
    xtick=data,
    width=45mm,
    height=26mm,
    ymin=0,
    ymax=50,
    axis y line*=none,
    axis x line*=none,
    xtick={1,2,3,4,5, 6},
    xticklabel style   = {align=center},
    xticklabels = {1600, 800, 400, 200, 100, 50},
    ytick={0, 10, 20, 30, 40, 50},
    yticklabels={0, 10, 20, 30, 40, 50},
    xlabel=No. cell executions,
    xlabel style={yshift = 2.5ex},
    ylabel style={yshift=-4ex},
    xmin = 0,
    xmax = 49,
    xtick = {0, 10, 20, 30, 40},
    xticklabels = {0, 10, 20, 30, 40},
    tick label style={font=\footnotesize},
    legend style={
        at={(-0.2,1.1)},anchor=south west,column sep=2pt,
        draw=black,fill=white,
        /tikz/every even column/.append style={column sep=5pt},
        inner ysep=0.5pt,
        font=\scriptsize,
    },
    legend cell align={left},
    legend columns=4,
    label style={font=\footnotesize},
    ylabel={No. of vars},
    ymajorgrids,
]

\addplot[GreenColor, mark = x, mark size=0.75pt, opacity = 0.7, densely dashed]
table[x=cellnum,y=total, densely dashed] {sections/data/appendix_profiling_var_count.txt};
\addlegendentry{Total data in session state}
\addplot[HeuristicColor, mark = *, mark size=0.75pt, opacity = 0.7]
table[x=cellnum,y=profiled] {sections/data/appendix_profiling_var_count.txt};
\addlegendentry{Data accessed per cell}



\end{axis}
\end{tikzpicture}
\vspace{-6.5mm}
\caption{Accessed variables}
\label{fig:background_access_vars}
\end{subfigure}
\begin{subfigure}[b]{0.48\linewidth}
\begin{tikzpicture}

\begin{axis}[
    xtick=data,
    width=45mm,
    height=26mm,
    ymin=0,
    ymax=2000000000,
    axis y line*=none,
    axis x line*=none,
    xtick={1,2,3,4,5, 6},
    xticklabel style   = {align=center},
    xticklabels = {1600, 800, 400, 200, 100, 50},
    ytick={0, 500000000, 1000000000, 1500000000, 2000000000},
    yticklabels={0, 0.5, 1, 1.5, 2},
    xlabel=No. cell executions,
    xlabel style={yshift = 2.5ex},
    label style={font=\footnotesize},
    ylabel style={yshift=-4ex, xshift=-0.5ex,font=\scriptsize},
    xmin = 0,
    xmax = 49,
    xtick = {0, 10, 20, 30, 40},
    xticklabels = {0, 10, 20, 30, 40},
    tick label style={font=\footnotesize},
    legend style={
        at={(-0.2,1.1)},anchor=south west,column sep=2pt,
        draw=black,fill=white,
        /tikz/every even column/.append style={column sep=5pt},
        font=\scriptsize,
    },
    legend cell align={left},
    legend columns=4,
    ylabel={Total size (GB)},
    ymajorgrids,
]

\addplot[GreenColor, mark = x, mark size=0.75pt, opacity = 0.7, densely dashed]
table[x=cellnum,y=total] {sections/data/appendix_profiling_var_size.txt};
\addplot[HeuristicColor, mark = *, mark size=0.75pt, opacity = 0.7]
table[x=cellnum,y=profiled] {sections/data/appendix_profiling_var_size.txt};



\end{axis}
\end{tikzpicture}
\label{fig:background_access_objs}
\vspace{-2.5mm}
\caption{Accessed objects}
\end{subfigure}
\hfill
\begin{subfigure}[b]{0.48\linewidth}
\begin{tikzpicture}

\begin{axis}[
    xtick=data,
    width=45mm,
    height=26mm,
    ymin=0,
    ymax=10,
    axis y line*=none,
    axis x line*=none,
    xtick={1,2,3,4,5, 6},
    xticklabel style   = {align=center},
    xticklabels = {1600, 800, 400, 200, 100, 50},
    ytick={2, 4, 6, 8, 10},
    yticklabels={2, 4, 6, 8, 10},
    xlabel=No. cell executions,
    xlabel style={yshift = 2.5ex},
    ylabel style={yshift=-4ex},
    xmin = 0,
    xmax = 49,
    xtick = {0, 10, 20, 30, 40},
    xticklabels = {0, 10, 20, 30, 40},
    tick label style={font=\footnotesize},
    legend style={
        at={(-0.2,1.1)},anchor=south west,column sep=2pt,
        draw=black,fill=white,
        /tikz/every even column/.append style={column sep=5pt},
        inner ysep=0.5pt,
        font=\scriptsize,
    },
    legend cell align={left},
    legend columns=4,
    label style={font=\footnotesize},
    ylabel={No. of vars},
    ymajorgrids,
]

\addplot[GreenColor, mark = x, mark size=0.75pt, opacity = 0.7, densely dashed]
table[x=cellnum,y=modify, densely dashed] {sections/data/appendix_create_modify_count.txt};
\addlegendentry{Modified data per cell}
\addplot[HeuristicColor, mark = *, mark size=0.75pt, opacity = 0.7]
table[x=cellnum,y=create] {sections/data/appendix_create_modify_count.txt};
\addlegendentry{Created data per cell}



\end{axis}
\end{tikzpicture}

\vspace{-2.5mm}
\caption{Modified vs. created variables}
\label{fig:background_append_update_vars}
\end{subfigure}
\begin{subfigure}[b]{0.48\linewidth}
\begin{tikzpicture}

\begin{axis}[
    xtick=data,
    width=45mm,
    height=26mm,
    ymin=0,
    ymax=600000000,
    axis y line*=none,
    axis x line*=none,
    xtick={1,2,3,4,5, 6},
    xticklabel style   = {align=center},
    xticklabels = {1600, 800, 400, 200, 100, 50},
    ytick={0, 200000000, 400000000, 600000000},
    yticklabels={0, 0.2, 0.4, 0.6},
    xlabel=No. cell executions,
    xlabel style={yshift = 2.5ex},
    label style={font=\footnotesize},
    ylabel style={yshift=-4ex, xshift=-0.5ex,font=\scriptsize},
    xmin = 0,
    xmax = 49,
    xtick = {0, 10, 20, 30, 40},
    xticklabels = {0, 10, 20, 30, 40},
    tick label style={font=\footnotesize},
    legend style={
        at={(-0.2,1.1)},anchor=south west,column sep=2pt,
        draw=black,fill=white,
        /tikz/every even column/.append style={column sep=5pt},
        font=\scriptsize,
    },
    legend cell align={left},
    legend columns=4,
    ylabel={Total size (GB)},
    ymajorgrids,
]

\addplot[GreenColor, mark = x, mark size=0.75pt, opacity = 0.7, densely dashed]
table[x=cellnum,y=modify] {sections/data/appendix_create_modify_size.txt};
\addplot[HeuristicColor, mark = *, mark size=0.75pt, opacity = 0.7]
table[x=cellnum,y=create] {sections/data/appendix_create_modify_size.txt};



\end{axis}
\end{tikzpicture}
\label{fig:background_append_update_objs}
\vspace{-2.5mm}
\caption{Modified vs. created objects}
\end{subfigure}
\vspace{-4mm}
\caption{Pattern of the \textit{TPS} notebook~\cite{tpsmay}. Contrast to the pattern of the in-progress \textit{Sklearn}~\cite{sklearntweet} notebook shown in \cref{fig:background_workload_characteristics}; despite \textit{TPS} being final and cleaned, the common patterns of (1) incremental executions and (2) balancing of data creation and modification are shared.
}
\label{fig:appendix_workload_characteristics}
\vspace{-3mm}
\end{figure}
\end{document}